\documentclass[letter]{article}
\pdfoutput=1
\usepackage{jheppub}
\usepackage{amsmath,amsfonts,relsize}
\usepackage[nameinlink,capitalize]{cleveref} 

\usepackage{graphicx}
\usepackage{dcolumn}
\usepackage[mathlines]{lineno}
\usepackage[caption=false]{subfig}
\usepackage{diagbox}
\usepackage{tabularx}
\usepackage{bm,color,xcolor}

\usepackage{floatrow}
\usepackage{comment}

\usepackage[rightcaption]{sidecap}
\sidecaptionvpos{figure}{c}
 
\newfloatcommand{capbtabbox}{table}[][\FBwidth]

\usepackage{enumitem}

\usepackage{xspace}

\usepackage{slashed,physics} 
\usepackage{float}
\usepackage{multirow}
\usepackage{mathtools}
\usepackage{appendix}
\usepackage[colorlinks=true
,urlcolor=blue
,anchorcolor=blue
,citecolor=blue 
,filecolor=blue
,linkcolor=blue
,menucolor=blue
,linktocpage=true
,pdfproducer=medialab
,pdfa=true
]{hyperref}
\usepackage{bm,color,color,soul}
\definecolor{UOgreen}{HTML}{007030}
\definecolor{matOrange}{HTML}{FF7F0E}
\definecolor{SNOLAB}{HTML}{2E88D1}
\definecolor{SUPL}{HTML}{D1772E}
\definecolor{grey}{HTML}{708090}

\usepackage{lineno}

\allowdisplaybreaks

\usepackage[normalem]{ulem}

\usepackage{subcaption}

\renewcommand{\vec}[1]{\bm{#1}}
\newcommand{\vmin}{v_{\rm min}}

\newcommand{\rhoDM}{\rho_{\rm \chi}}
\newcommand{\mDM}{m_{\rm \chi}}

\newcommand{\NE}{$e^-$}
\newcommand{\sige}{\overline\sigma_e}
\newcommand{\FDM}{F_{\rm DM}}

\newcommand{\beq}{\begin{equation}}
\newcommand{\eeq}{\end{equation}}
\newcommand{\bea}{\begin{eqnarray}}
\newcommand{\eea}{\end{eqnarray}}

\usepackage[dvipsnames]{xcolor}

\graphicspath{{figs/}}

\begin{document}


\title{Earth-Scattering Induced Modulation in Low-Threshold Dark Matter Experiments}

\author[a]{Xavier Bertou}
\emailAdd{bertou@ijclab.in2p3.fr}
\affiliation[a]{\normalsize\it Laboratoire de Physique des 2 infinis Irène Joliot-Curie – IJCLab (CNRS/IN2P3), Orsay, France}

\author[b]{, Ansh Desai}
\emailAdd{adesai@uoregon.edu}
\affiliation[b]{\normalsize\it Department of Physics and Institute for Fundamental Science,
University of Oregon, Eugene, Oregon 97403, USA}

\author[c]{, Timon Emken}
\emailAdd{timon.emken@fysik.su.se}
\affiliation[c]{\normalsize\it The Oskar Klein Centre, Department of Physics, Stockholm University, AlbaNova, SE-10691 Stockholm, Sweden}

\author[d]{, Rouven Essig}
\emailAdd{rouven.essig@stonybrook.edu}
\affiliation[d]{\normalsize\it C.N.~Yang Institute for Theoretical Physics, Stony Brook University, Stony Brook, NY 11794, USA}

\author[e]{, Tomer Volansky}
\emailAdd{tomerv@post.tau.ac.il}
\affiliation[e]{\normalsize\it 
 School of Physics and Astronomy,   
 Tel-Aviv University, Tel-Aviv 69978, Israel}
 
\author[b]{, Tien-Tien Yu}
\emailAdd{tientien@uoregon.edu}

\abstract{Dark matter particles with sufficiently large interactions with ordinary matter can scatter in the Earth before reaching and scattering in a detector.  This induces a modulation in the signal rate with a period of one sidereal day.  We calculate this modulation for sub-GeV dark matter particles that interact either with a heavy or an ultralight dark-photon mediator and investigate the resulting signal in low-threshold detectors consisting of silicon, xenon, or argon targets.  The scattering in the Earth is dominated by dark matter scatters off nuclei, while the signal in the detector is easiest to observe from dark matter scattering off electrons. We investigate the properties of the modulation signal and provide projections of the sensitivity of future experiments.  We find that a search for a modulation signal can probe new regions of parameter space near the energy thresholds of current experiments, where the data are typically dominated by backgrounds.}

\maketitle

\section{Introduction}
There is significant evidence supporting the existence of an invisible, non-baryonic form of matter known as dark matter (DM), but despite the breadth of experimental searches, its particle nature remains unknown. A promising avenue of research to tackle this problem is the direct detection of galactic DM, a field that has made significant advancements over the last decade. Of the zoo of DM candidates, a particularly interesting region of parameter space is light DM ($\mDM<1$ GeV), which is strongly motivated and can produce a measurable signal through DM-electron scattering~\cite{Essig:2011nj,essig2023snowmass2021cosmicfrontierlandscape}.  
Such scatters produce one to a few electrons in semiconductor and noble liquid targets. 
In order to help distinguish a DM signal from backgrounds, especially in the low charge bins, it is helpful to consider distinctive properties of the DM signal, such as a modulation in the scattering rate.  

It is well known that a DM signal is expected to modulate annually due to the orbit of the Earth around the Sun~\cite{drukier_1986}.  Additionally however, the signal can modulate with the time period of a sidereal day due to effects such as gravitational focusing~\cite{nielsen_grav_focusing,Sikivie_2002,Alenazi_2006}, or the daily fluctuation of Earth's rotational velocity with respect to the DM halo~\cite{nielsen_grav_focusing}. 
Moreover, if the DM interaction with ordinary matter is sufficiently large, interactions with the bulk matter of the Earth can distort the velocity distributions of the DM particles as they travel through the Earth~\cite{COLLAR1992181,PhysRevD.47.5238,Hasenbalg_1997,Emken_2017}, which can greatly enhance or attenuate the flux seen at a detector. Consequently, the location of the detector on Earth relative to the DM wind will result in considerably different predictions for the expected signal, and will modulate over the course of Earth's rotation. This latter form of daily modulation is the focus of this paper. 

In Section~\ref{sec:formalism}, we briefly review DM-electron scattering and note that the integrated velocity distribution used to calculate the event rates can be modified to account for DM-Earth scattering. In Section~\ref{sec:earth}, we detail the DM benchmark model that will be the focus of this paper: DM interacting with a dark photon. In  this case, the DM-Earth scattering is dominated by DM-nucleus scattering, while the signal at the detector will be generated predominantly through DM-electron scattering.  We introduce the two tools we use to simulate DM-nucleus scattering in the Earth. The first is {\tt DaMaSCUS}~\cite{Emken_2017}, which provides a full 3D Monte Carlo simulation of DM particles traversing the Earth for a given location and time of year. The second is {\tt Verne}~\cite{Kavanagh_2018}, which makes an analytic approximation that greatly simplifies the calculation, but is expected to lose accuracy for cross-sections where DM undergoes multiple scatters within the Earth.  {\tt DaMaSCUS} is computationally intensive, particularly for large cross-sections, but is more accurate than {\tt Verne}.  We compare these two tools and study in detail the effect of daily modulation in silicon, xenon, and argon in Section~\ref{sec:results}. We also study the expected discovery potential in silicon for various exposures and see that for the lower charge bins, experiments can significantly improve their sensitivity by performing a daily modulation search. Discovery potential in xenon and argon is discussed in Appendix~\ref{app:noble}. Finally, we summarize our findings in Section~\ref{sec:conclusion}. In the main part of the paper we focus on the northern hemisphere, but Appendix~\ref{app:south} contains additional figures for the southern hemisphere.

\section{Dark Matter-Electron Scattering}
\label{sec:formalism}
\subsection{Formalism}
\label{sec:rates}
For an isotropic DM velocity distribution, the general expression for the differential DM-electron scattering rate is given by~\cite{Essig:2011nj,Essig:2015cda}
\beq
\frac{dR}{d \ln E_{e}} = \frac{\rhoDM}{m_{\chi}}\frac{\sige}{8\mu^{2}_{\chi e}}\int  \mathrm{~d}q \ q|\FDM(q)|^{2}  \vert  f_{\rm res}(E_e, q)\vert^{2} \eta \left(\vmin\right)\, ,
\label{eq:ionization}
\eeq
where $E_e$ is the energy of the outgoing electron, $m_\chi$ is the DM mass, $q$ is the momentum transfer of the DM-electron interaction, $\rhoDM$ is the local DM density, and $\mu_{\chi e}$ is the reduced-mass of the DM-electron system. We parameterize the underlying DM-electron interaction using the usual definitions~\cite{Essig:2011nj}
\begin{align}
    \overline{|{\cal M}(\vec q)|^2}&\equiv \overline{|{\cal M}(q_{\rm ref})|^2}\times |F_{\rm DM}(q)|^2\\
    \overline\sigma_e&\equiv\frac{\mu_{\chi e}^2 \overline{|{\cal M}(q_{\rm ref})|^2}}{16\pi m_\chi^2m_e^2}\, ,
\end{align}
with  $\overline{|{\cal M}|^2}$ being the spin-averaged matrix-element squared, and we set the reference momentum $q_{\rm ref}=\alpha m_e$. The momentum dependence of the interaction is encoded in $ F_{\rm DM}(q)=(q_{\rm ref}/q)^n$, where $n=0,1,2$ corresponds to a contact-interaction, electric dipole coupling, or long-range interaction, respectively.

The material-dependent form factor, $f_{\rm res}(E_e, q)$ encapsulates the 
system's response 
for an electron excitation with momentum $q$ and energy $E_e$. For an atomic-target, we have
\beq
|f_{\rm res}(E_e,q)|^2 \equiv |f_{nl}^{\rm ion}(E_e,q)|^2\, ,
\eeq
where $(n,l)$ are the quantum numbers for the initial-state bound electron and $k=\sqrt{2 m_e E_e}$ is the momentum of the final-state electron. More specifically, the atomic form factor  is given by 
\bea
|f^{\rm ion}_{nl}(k,q)|^2=\frac{4k^{3}}{(2\pi)^3}\sum_{l' L}(2l+1)(2l'+1)(2L+1)&&\nonumber\\
\times\left[\begin{matrix} l&l'&L\\0&0&0\end{matrix}\right]^2
 \left|\int r^2\mathrm{~d}r R_{kl'}(r)R_{nl}(r)j_L(qr) \right|^2,&&
\eea
where $[\cdots]$ is the Wigner 3-$j$ symbol and $j_L$ are the spherical Bessel functions. In this work, we take the incoming wavefunctions $R_{nl}$ to be Roothaan-Hartree-Fock (RHF) ground-state wave functions, while the outgoing wavefunctions $R_{kl'}$ are solutions to the Schr\"odinger equation with a hydrogenic potential $-Z_{\rm eff}/r$~\cite{Essig:2012yx,Essig:2017kqs, Catena:2019gfa}.

For a crystal-target, we have,
\beq
|f_{\rm res}(E_e, q)|^2 \equiv \frac{8\alpha_{\rm EM} m_e^2 E_e}{q^3}\times |f_{\rm crystal}(E_e,q)|^2\, ,
\eeq
where $\alpha_{\rm EM}$ is the fine-structure constant, $m_e$ is the electron mass, and $f_{\rm crystal}(E_e,q)$ is the dimensionless crystal form factor as defined in~\cite{Essig:2015cda}. There are a variety of techniques for calculating $f_{\rm crystal}$ in solid-state materials using either single-particle wavefunctions extracted from density functional theory (DFT), as found in {\tt QEDark}~\cite{Essig:2015cda}, {\tt QEDark-EFT}~\cite{Catena:2021qsr}, {\tt EXCEED-DM}~\cite{Griffin:2021znd}, and {\tt QCDark}~\cite{Dreyer:2023ovn}, or the energy-loss function (ELF), which encodes the many-body response function as found in {\tt DarkELF}~\cite{Knapen:2021bwg}  (see also~\cite{Hochberg:2021pkt}). Note that the latter applies for scenarios in which the DM-electron interactions are spin-independent and couple to the electron density, and furthermore accounts for the charge-screening effect in the detector. For $m_\chi\lesssim 10$ MeV, this screening effect reduces the rate by ${\cal O}(1)$. In this work, we use the crystal form factor from {\tt QCDark}, which includes a screening factor function~\cite{PhysRevB.47.9892}
\beq
\epsilon(q,\omega)=1+\left[\frac{1}{\epsilon_0-1}+\alpha\left(\frac{q}{q_{\rm TF}}\right)^2+\frac{q^4}{4 m_e^2\omega_p^2}-\left(\frac{\omega}{\omega_p}\right)^2\right]^{-1}\, ,
\eeq
where in silicon $\epsilon_0=\epsilon(0,0)=11.3$ is the static dielectric constant, $\alpha=1.563$ is a fitting parameter, $\omega_p=16.6$~eV is the plasma frequency, and $q_{\rm TF}=4.13$~keV is the Thomas-Fermi momentum~\cite{Griffin:2021znd}.
We note that these differences between {\tt QCDark} and other software such as {\tt DarkELF} affect the scattering rate in the detector but not the scattering rate in the Earth. 

The DM halo velocity distribution in Eq.~\eqref{eq:ionization} is encoded in
\begin{equation}
\label{eq:eta}
    \eta(v_{\rm min})=\int_{v_{\rm min}}\frac{\mathrm{~d}^3 v}{v}f_{\chi}(\vec v)\, .
\end{equation}
Typically, the velocity distribution $f_{\chi}(\vec v)$ is taken to be a time-independent, isotropic Maxwell-Boltzmann distribution, leading to what is known as the Standard Halo Model (SHM)~\cite{Lewin:1995rx,Green_2017}. However, as we will show in Section \ref{sec:earth}, effects from DM-Earth scattering will modify the distribution and lead to interesting daily modulation effects. These effects can be encoded in a time-dependent velocity distribution, $f_{\chi}(\vec v,t)$, which result in a time-dependent DM-electron scattering rate,
\beq
\label{eq:rate}
R(t)=\frac{\rhoDM}{m_{\chi}}\frac{\sige}{8\mu^{2}_{\chi e}}\int \frac{\mathrm{~d}E_e}{E_e} \int  \mathrm{~d}q \ q|\FDM(q)|^{2}  \vert  f_{\rm res}(E_e, q)\vert^{2} \eta \left(\vmin,t\right)\, .
\eeq
For this analysis, we use halo parameters recommended in~\cite{Baxter:2021pqo}, with $v_0 = 238$ km/s, $v_{\mathrm{esc}} = 544$ km/s, and $\rho_\chi = 0.3$ GeV/cm$^3$ (see also~\cite{Catena:2009mf, Salucci:2010qr, Read:2014qva, deSalas:2020hbh,Lim:2023lss}). 

\subsection{Dark Photon Mediator}
For the purposes of this work, we will consider a fermionic DM candidate,\footnote{Our results also hold for a scalar DM candidate.} $\chi$, that interacts with the SM through a dark photon $A'$ with the Lagrangian 
\begin{subequations}\label{eq: lagrangian dark photon}
\begin{align}
    \mathcal{L}_{D}&=\bar{\chi}\left(i \gamma^{\mu} D_{\mu}-m_{\chi}\right) \chi+\frac{1}{4} F_{\mu \nu}^{\prime} F^{\prime \mu \nu}
    +m_{A^{\prime}}^{2} A_{\mu}^{\prime} A^{\prime \mu}+\frac{\varepsilon}{2} F_{\mu \nu} F^{\prime \mu \nu}\, , 
\end{align}
\end{subequations}
where $m_{A'}$ is the mass of the dark photon, $D_\mu=\partial_\mu-i g_D A^\prime_\mu $, with $g_D$ the dark gauge coupling.
The corresponding scattering cross-sections for DM-nucleus ($\sigma_N$) and DM-electron ($\sigma_e$) interactions are then 
    \begin{align}
\frac{\mathrm{d} \sigma_{N}}{\mathrm{~d} q^{2}} &=\frac{\bar{\sigma}_{p}}{4 \mu_{\chi p}^{2} v^{2}} F_{\mathrm{DM}}(q)^{2} F_{N}(q)^{2} Z^{2} \, ,\\
\frac{\mathrm{d} \sigma_{e}}{\mathrm{~d} q^{2}} &=\frac{\bar{\sigma}_{e}}{4 \mu_{\chi e}^{2} v^{2}} F_{\mathrm{DM}}(q)^{2}\, ,
\end{align}
where $Z$ is the atomic number of the nuclear target and $v$ is the relative velocity between the DM and the target.  $F_N(q)$ is the nuclear form factor, taken to be a Helm form factor~\cite{Lewin:1995rx}, and 
\begin{align}
    \bar{\sigma}_{p} =\frac{\mu_{\chi p}^{2}}{\mu_{\chi e}^{2}}  \bar{\sigma}_{e} = \frac{16 \pi \alpha \alpha_{D} \epsilon^{2} \mu_{\chi p}^{2}}{\left(q_{\mathrm{ref}}^{2}+m_{A^{\prime}}^{2}\right)^{2}}\,,
\end{align}
with $\alpha_D=g_D^2/4\pi$.
Note that the relationship between the DM-electron and DM-proton scattering cross-section would generically differ for other DM models. 
To account for charge screening in the DM-nucleus scattering, we make the replacement as in~\cite{PhysRev.83.252, RevModPhys.46.815}. 
\begin{align}
    Z \rightarrow Z_{\text {eff }}=F_{A}(q) \times Z\,,
\end{align}
where
\begin{align}
F_{A}(q)=\frac{a^{2} q^{2}}{1+a^{2} q^{2}}
\end{align}
and
\begin{align}
a=\frac{1}{4}\left(\frac{9 \pi^{2}}{2 Z}\right)^{1 / 3}\, ~ a_{0} \approx \frac{0.89}{Z^{1 / 3}} a_{0}\, 
\end{align}

\section{Dark Matter-Earth Scattering}\label{sec:earth}

If the interaction strength between halo DM particles and nuclei is sufficiently large, scatterings with terrestrial nuclei occur frequently enough to alter the distribution of the underground flux~\cite{PhysRevD.47.5238}. In this work, we demonstrate that such Earth-induced scatterings occur in the parameter space accessible to current low-threshold experiments targeting DM-electron interactions. These scatterings deflect and decelerate DM particles, thereby modifying both their spatial distribution and velocity spectrum.
As a result, the flux of DM particles incident on an underground detector becomes location-dependent, varying with the detector's orientation relative to the DM wind. Subsurface scatterings can either attenuate or enhance the local DM flux. Consequently, the Earth's rotation induces a daily modulation in the signal rate of direct-detection experiments.\footnote{We note that while we only consider halo DM in this work, solar-reflected DM (see~\cite{An_2018,An_2021,Emken_2022,Emken:2024nox}) is also expected to modulate, though it has a markedly different velocity distribution and modulates with a different phase and period.} While in our model DM scatters off of nuclei and electrons in the Earth, the largest effect on the resulting DM velocity distributions arises from the DM-nucleus scatterings. This introduces a degree of model dependence into our analysis; we adopt the dark photon model as a representative benchmark.

\begin{figure*}[!t]
    \centering
 \includegraphics[width=0.75\textwidth]{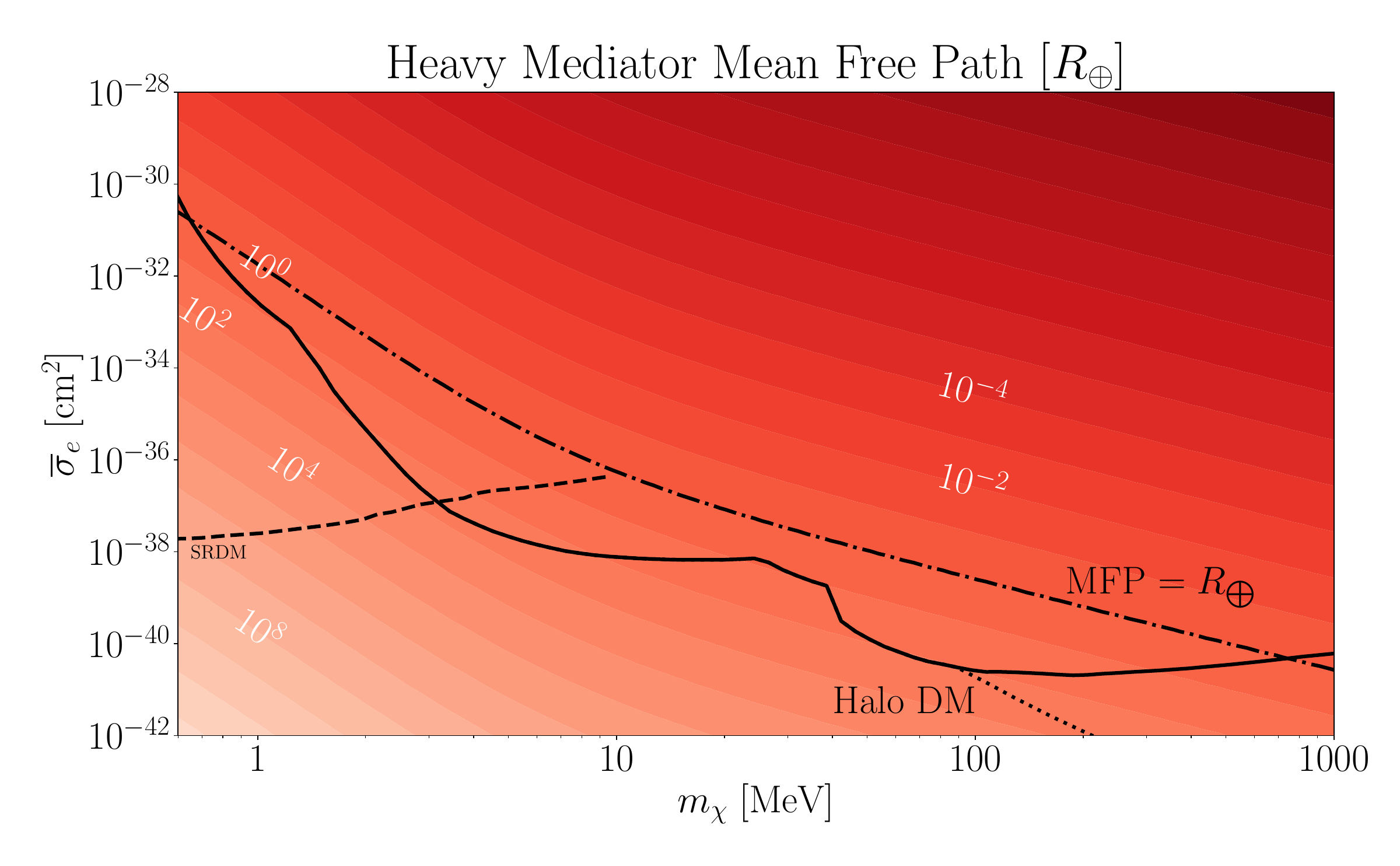}
     \includegraphics[width=0.75\textwidth]{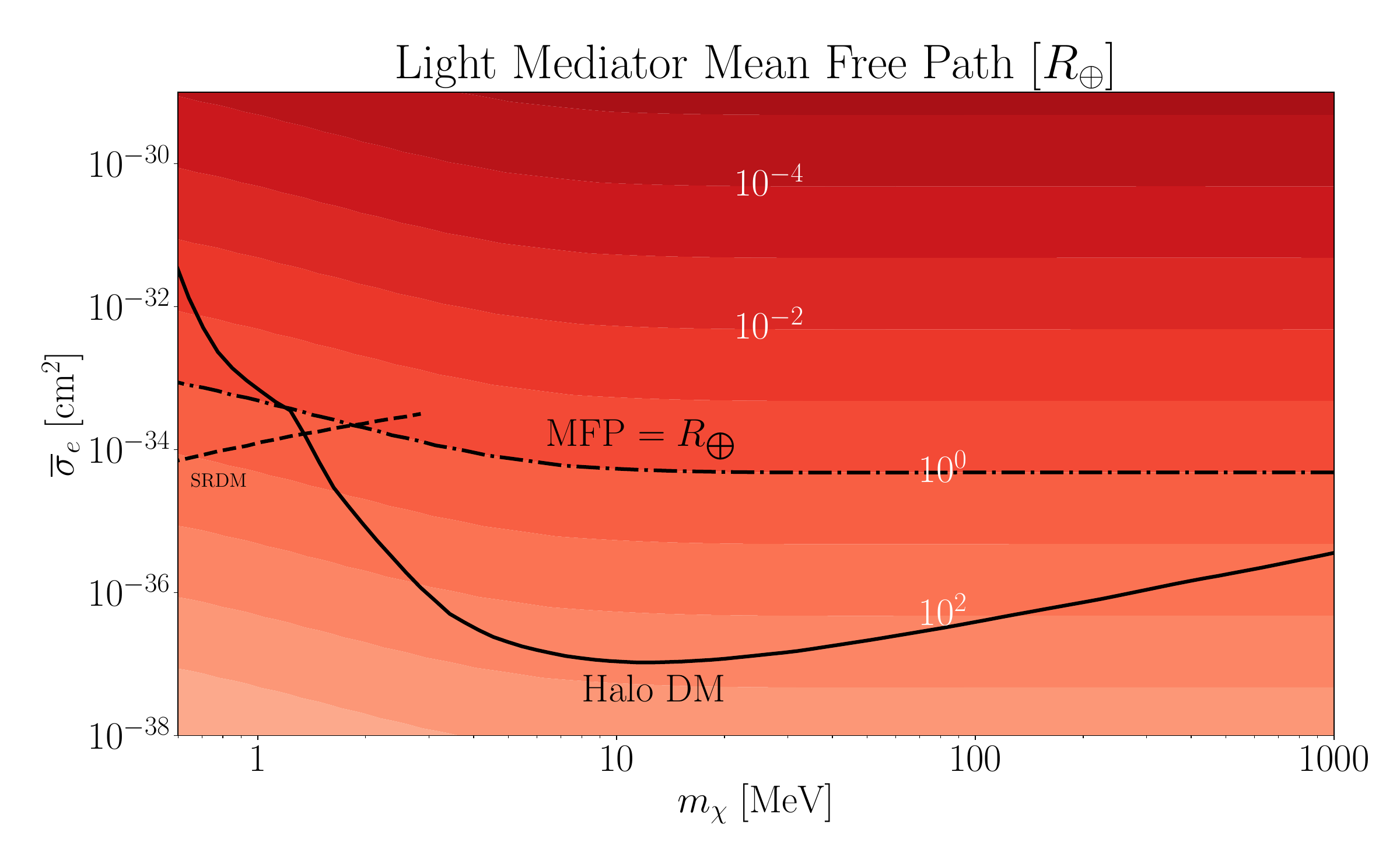}   
    \caption{Contours of the mean free path through the mantle $\lambda$ in units of Earth radius ($R_\oplus$) in the DM mass $m_\chi$ and DM-electron scattering cross-section $\overline\sigma_e$ parameter space, for a heavy mediator ({\bf top}) and light mediator ({\bf bottom}). 
For the light mediator, the halo constraints ({\bf solid black}) are taken from~\cite{SENSEI:2024yyt,damicmcollaboration2025probingbenchmarkmodelshiddensector,DamicModArnquist_2024} while for the heavy mediator,  constraints are combined using~\cite{damicmcollaboration2025probingbenchmarkmodelshiddensector,DamicModArnquist_2024,SENSEI:2024yyt,PandaXTLi_2023,XENON:2024znc,DarkSide:2022knj}. Constraints from the Migdal effect from~\cite{PandaXTLi_2023} are shown in {\bf dotted black}.  The {\bf dashed black} line shows the excluded region which arises from the solar reflected constraint from DM~\cite{XENON:2024znc}. Along the dash-dotted line, the mean free path equals $R_\oplus$.}
    \label{fig:mfp}
\end{figure*}

In this section, we summarize how DM-nucleus scattering is simulated within the Earth.
For a more detailed description, we refer the reader to~\cite{Emken:2017qmp,Emken:2019hgy}. 
The scattering frequency is predominantly determined by the DM mean free path,
\begin{align}
    \lambda^{-1} = \sum_i \lambda_i^{-1}\equiv \sum_i n_i(r)\sigma_i = \sum_i f_i(r) \frac{\rho_\oplus(r)}{m_i}\sigma_i\, ,\label{eq: mean free path}
\end{align}
where the index $i$ runs over the nuclear isotopes present in the Earth, $n_i(r) = f_i(r) \frac{\rho_\oplus(r)}{m_i}$ denotes the corresponding number density, where $f_i$ is the fractional density of the $i$th isotope of mass $m_i$, and $\rho_\oplus(r)$ is the mass density profile of the Earth.
The density profile we use is part of the Preliminary Reference Earth Model (PREM)~\cite{Dziewonski:1981xy}, and the fractional densities for the Earth core and mantle are given in~\cite{McDonough:2003}. The exact mean free path will depend on the trajectory of a DM particle, as it determines which of the Earth's layers it passes through. We show an example of the mean free path through the mantle for both light and heavy mediators in Fig.~\ref{fig:mfp}.

The total cross-section~$\sigma_i$ is given by
\begin{align}
\sigma_{i}&=\int_{0}^{q_{\max }^{2}} \mathrm{~d} q^{2} \frac{\mathrm{d} \sigma_{i}}{\mathrm{~d} q^{2}}\,,
\end{align}
which, depending on the mass of the mediator and charge screening, evaluates to
\begin{align}
\sigma_{i}&=\bar{\sigma}_{p}\left(\frac{\mu_{\chi i}}{\mu_{\chi p}}\right)^{2} Z^{2}\\
&\times \begin{cases}
    1& \text { for } F_{\mathrm{DM}}(q)=1, \text{ without screening}\\
    \left[1+\frac{1}{1+a^{2} q_{\max }^{2}}-\frac{2}{a^{2} q_{\max }^{2}} \log \left(1+a^{2} q_{\max }^{2}\right)\right], & \text { for } F_{\mathrm{DM}}(q)=1, \text{ with screening} \\
    \frac{a^{4} q_{\mathrm{ref}}^{4}}{\left(1+a^{2} q_{\max }^{2}\right)}, & \text { for } F_{\mathrm{DM}}(q)=\left(\frac{q_{\text {ref }}}{q}\right)^{2}\text{ with screening} .\end{cases}\nonumber
\end{align}
The maximal momentum transfer in a scattering between a DM particle and nucleus~$i$ is $q_{\max }=2 \mu_{\chi i} v$, where $\mu_{\chi i}$ is the reduced mass.



\begin{figure*}[t!]
    \centering
\includegraphics[width=0.65\textwidth]{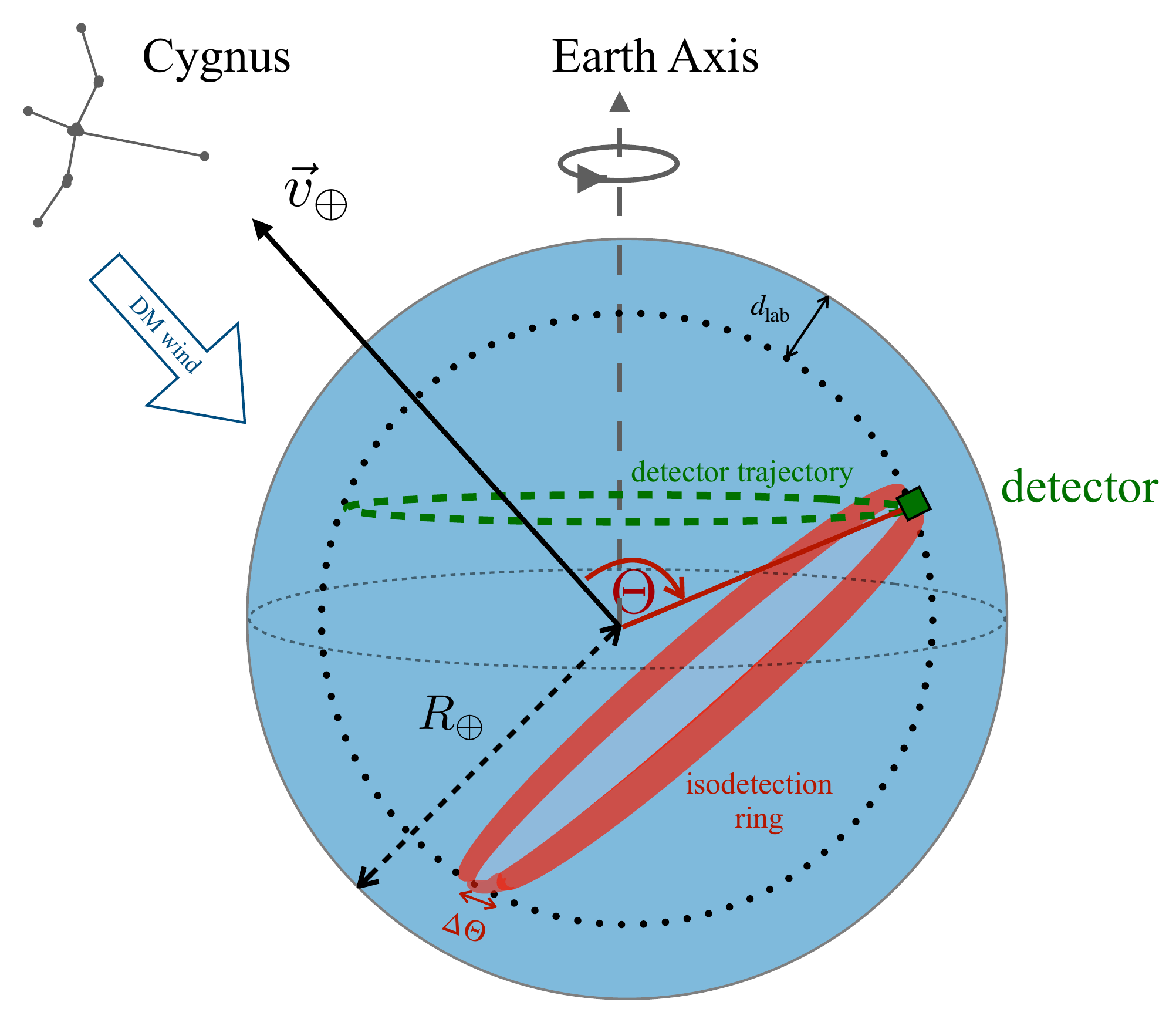}
    \caption{An example of an isodetection ring with a finite $\Delta \Theta$. The Earth's velocity, $\vec v_\oplus$, points in the direction of the Cygnus constellation.}
    \label{fig:isoring}
\end{figure*}

The key parameter governing the modulation is the isodetection angle (isoangle, denoted by $\Theta$), defined as the angle between the Earth's velocity vector and the position of the detector as measured from the center of the Earth (see Fig.~\ref{fig:isoring}). To illustrate this, consider a spherical surface within the Earth of radius $r = R_\oplus-d_{\rm lab}$, where $d_{\rm lab}$ is the depth of the detector below the surface. A detector at fixed geographic coordinates traces a nontrivial path through these isoangles due to Earth's rotation and orbital motion. The dynamics of this motion and its implications are discussed in detail in~\cite{Emken_2017}. The detector's position, expressed in terms of the isoangle $\Theta$, is given by,
\begin{align}
    \Theta(t) = \arccos\left({\frac{\vec{v}_{\oplus} \cdot \vec{x}_{\rm lab}(t)}{v_{\oplus} (R_\oplus - d_{\rm lab})}} \right)\,,
    \label{isodef}
\end{align}
where $\vec{x}_{\rm lab}$ is the galactic lab position vector and
\begin{align}
    \vec{v}_\oplus (t) = \vec{v}_r + \vec{v}_s  + \vec{v}_e (t)\, .
 \end{align}
Here $\vec{v}_r$ is the galactic rotation, $\vec{v}_s$ is the Sun's motion relative to nearby stars, and $\vec{v}_e(t)$ is the Earth's orbital velocity relative to the Sun. Each location on Earth has a certain isoangle range associated with it that modulates with the period of a sidereal day. A key feature is that, along a constant isoangle, the DM particle's velocity distribution and direct-detection event rates are constant.  In the Monte Carlo simulation, we divide up the Earth into 36 rings of size $\Delta\Theta=1^\circ$ (spaced by $5^\circ)$ , where $\Delta\Theta$ is sufficiently small such that the velocity distribution will be approximately constant over the surface of a single isodetection ring. 

\begin{figure*}[!t]
    \includegraphics[width=0.99\textwidth]{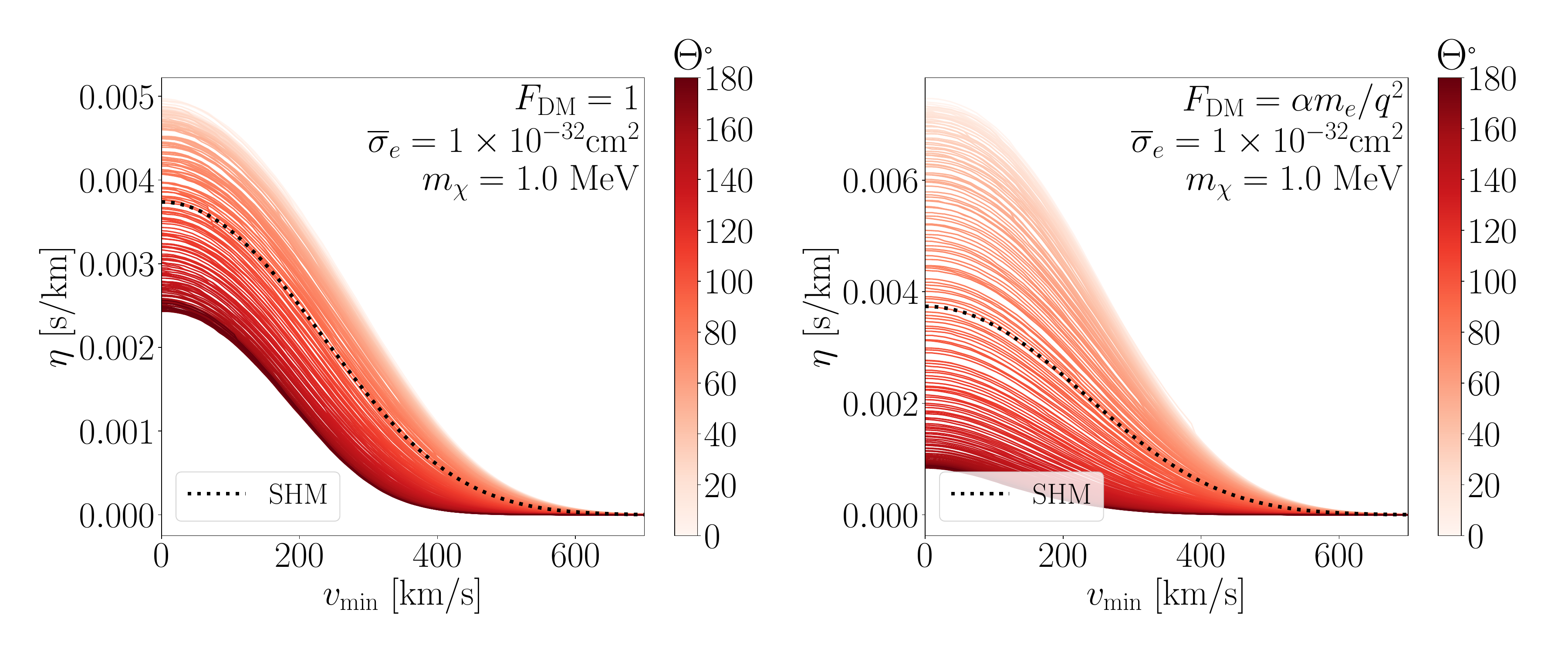}
    \caption{Examples of the shift in $\eta$ (defined in Eq.~(\ref{eq:eta})) as a function of isoangle $\Theta$ for DM interacting with a heavy mediator ({\bf left}) or a light mediator ({\bf right}), for a DM mass of $m_\chi=1$ MeV, and DM-electron scattering cross-section of $\sigma_e=10^{-32}$ cm$^2$. The unshifted $\eta$ for the SHM is shown as a {\bf black, dashed} line in both panels. These lines are calculated using {\tt DaMaSCUS}.}
    \label{fig:iso_eta}
\end{figure*}

For large cross-sections, it is necessary to perform Monte Carlo (MC) simulations of underground DM trajectories to account for multiple scatterings~\cite{PhysRevD.47.5238,Hasenbalg_1997,Emken:2017qmp}. 
For this study we make use of the publicly available {\tt DaMasSCUS} code to perform full 3D Monte-Carlo simulations of DM-nucleus scattering within the Earth for a given date (chosen where $v_E$ is average, which occurs on March 8th\footnote{The choice of year is arbitrary \cite{Baxter:2021pqo}.})~\cite{Emken2017_code}. We also note two new features of \texttt{DaMaSCUS}, which are part of a new public branch of the software. First, we add the possibility of simulating DM~particles interacting with nuclei via light mediators, and second we include interactions where the nuclear charge is screened on larger scales.  
An example of the distorted velocity distributions simulated with {\tt DaMaSCUS} as a function of isoangle can be found in Fig.~\ref{fig:iso_eta}. Certain points are out of reach for these MC simulations due to either computational complexity (too many scatters) or insufficient statistics (too few scatters). We note that much of the parameter space relevant to the former problem has been ruled out. To study the latter points, we make use of analytic approximations (described below).\footnote{MC simulation of DM trajectories have also been used to quantify the DM flux attenuation of the overburden of direct detection experiments~\cite{Zaharijas:2004jv,Emken_2017,Mahdawi:2017cxz,Emken:2018run,Mahdawi:2018euy,Emken:2019tni}.}

In the regime for which a single scattering occurs on average, daily modulations induced by Earth scatterings can be accurately described using analytic methods~\cite{Kavanagh:2016pyr,Emken:2021vmf}. This approximation holds when the DM mean free path is comparable to or exceeds the Earth's radius (see Fig.~\ref{fig:mfp}). The publicly available {\tt Verne} code implements such an analytic approach~\cite{Kavanagh_2018}, modeling DM particles as either traversing the Earth unimpeded or undergoing complete reflection upon a single scatter. This allows us to study daily modulation even for regions of parameter space that have a large mean free path compared to $R_{\bigoplus}$, which are out of reach for {\tt DaMaSCUS}. We emphasize that our analysis employs a newly updated version of \texttt{Verne}, distinct from earlier versions used in prior studies, such as~\cite{DamicModArnquist_2024}. This code takes into account the effects of atmospheric overburden; the parameter space for which this would be relevant is towards larger cross sections than shown in our figures. For certain combinations of larger masses ($m_\chi > 100$ MeV) and larger cross-sections ($\overline{\sigma}_e > 10^{-30}$) the flux can be completely attenuated, but for the masses and cross-sections that are the focus of this paper, this effect is negligible compared to the scattering cross-sections within the Earth.



To study the time-dependent variation in DM event rates, we consider fixed experimental latitudes. We select two representative underground laboratory sites, one in each hemisphere. For the Northern Hemisphere, we use SNOLAB (Sudbury, Ontario, Canada), a representative example given that most northern underground facilities that host direct-detection experiments such as Fermilab (USA), Canfranc (Spain), Modane (France), and Gran Sasso (Italy) are located at similar latitudes. Consequently, the expected modulation behavior is largely consistent across these sites. For the Southern Hemisphere, we choose the Stawell Underground Physics Laboratory (SUPL) in Victoria, Australia. This analysis is also relevant to the proposed Paarl Africa Underground Laboratory near Cape Town, South Africa~\cite{adam2023paarlafricaundergroundlaboratory}, which lies approximately $3^\circ$ further north than SUPL. 

\begin{figure*}[!t]
  \includegraphics[width=1.0\textwidth]{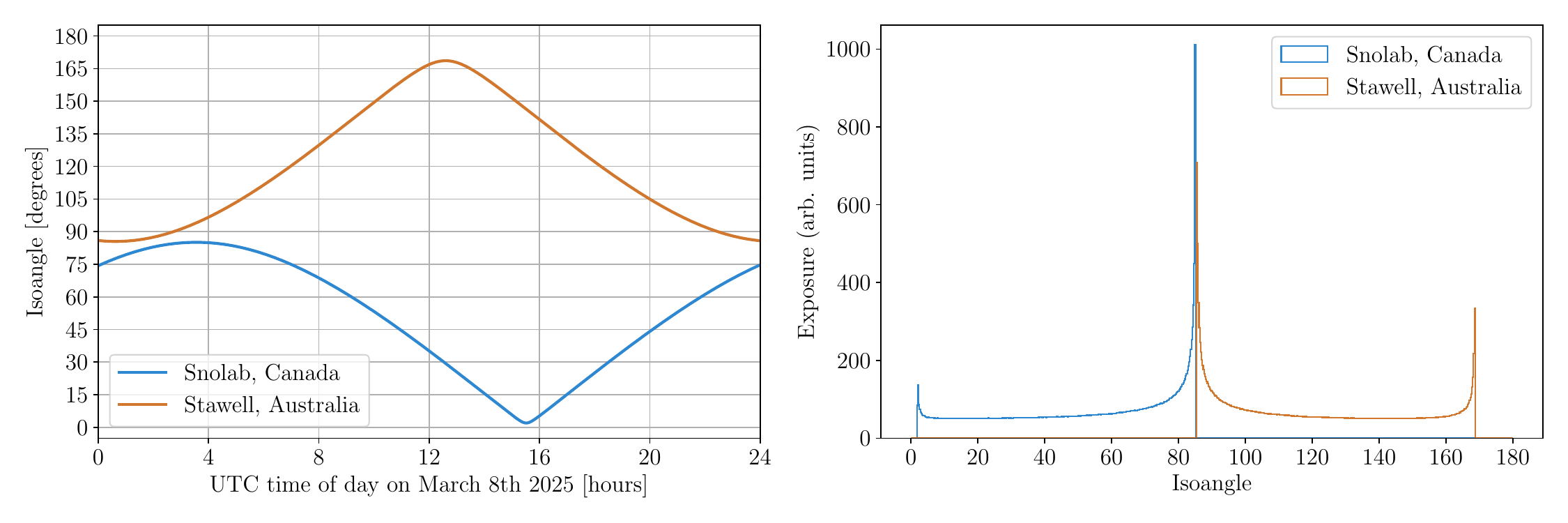}
   \caption{{\bf Left}: Isoangle as a function of time for a detector at SNOLAB ({\bf\color{SNOLAB}blue}) and SUPL ({\bf\color{SUPL} orange}) on March 8th, 2025). {\bf Right}: Differential exposure as a function of isoangle for the same sites.}
    \label{fig:isoangle vs time}
\end{figure*}
An illustrative comparison of isoangle variation over time for the two selected locations is shown in the left panel of Fig.~\ref{fig:isoangle vs time}. At SNOLAB (46$^\circ$N), during December, the detector traverses isoangles $\Theta \in [6^\circ, 81^\circ]$, whereas at SUPL (37$^\circ$S), it spans $\Theta \in [89^\circ, 165^\circ]$. These ranges shift by a degree or two annually, depending on the site's latitude. Notably, a detector at the geographic poles would exhibit an almost flat isoangle-time profile, while one at the equator would display a symmetric curve centered around the equinoxes. The right panel of Fig.~\ref{fig:isoangle vs time} presents the exposure as a function of isoangle, i.e., the time spent at each isoangle, with peaks corresponding to the turning points observed in the left panel.

\section{Results}
\label{sec:results}

\subsection{Modulation Rates}\label{sec:mod}

For silicon, the event rates in this study are computed using a modified version of {\tt QCDark}, incorporating dielectric screening as detailed in Section~\ref{sec:rates}. We replace the SHM velocity distribution with isoangle-dependent distributions, $f_\chi(\vec v,\Theta(t))$, derived from {\tt DaMaSCUS} and a modified version of {\tt Verne}, integrated according to Eq.~\eqref{eq:eta}. Consequently, the calculated rates are explicitly isoangle-dependent, as in Eq.~\eqref{eq:rate}. For silicon, we focus on the 1$e^-$ ionization bin, using the ionization probabilities from~\cite{Ramanathan_2020}.

For xenon and argon, we similarly adapt {\tt wimprates}~\cite{wimprates} to incorporate isoangle-dependent velocity distributions. Since argon is not included in the public release of {\tt wimprates}, we compute the relevant form factors following the formalism outlined in Section~\ref{sec:rates}. We consider the 1--4~\NE{} ionization bins for both targets. Secondary ionization is treated following~\cite{Essig:2012yx,Essig:2017kqs}, using $W = 13.8$ eV (19.5 eV), $f_R = 0$, and $f_e = 0.83$ for xenon (argon), where $W$ is the average energy to produce a single quanta, $f_R$ is the recombination probability, and $f_e$ is the electron yield fraction.

To analyze the modulation across a broad range of recoil energies and cross-sections, we fit the simulated event rates (from {\tt DaMaSCUS}), $R(\Theta)$, as a function of isoangle, and therefore time, using a hyperbolic tangent function of the form 
\begin{equation}\label{eq: hyptan fit}
   R(\Theta) = \frac{A}{2}\ \tanh{\left(\frac{\Theta-\Theta_0}{\Theta_S}\right)} + C\, ,
\end{equation}
where the parameters $A$, $\Theta_0$, $\Theta_S$, and $C$ correspond to the modulation amplitude, the transition angle, the slope of the transition, and a constant offset, respectively. These fits are performed across the full grid of simulated parameter points and provide a smooth functional description of the isoangle dependence of the rate. This allows for interpolation between discrete simulation points and enables the predictions of the rate at arbitrary combinations of energy and cross-section.

\begin{figure*}[t]

  \includegraphics[clip,width=0.88\columnwidth]{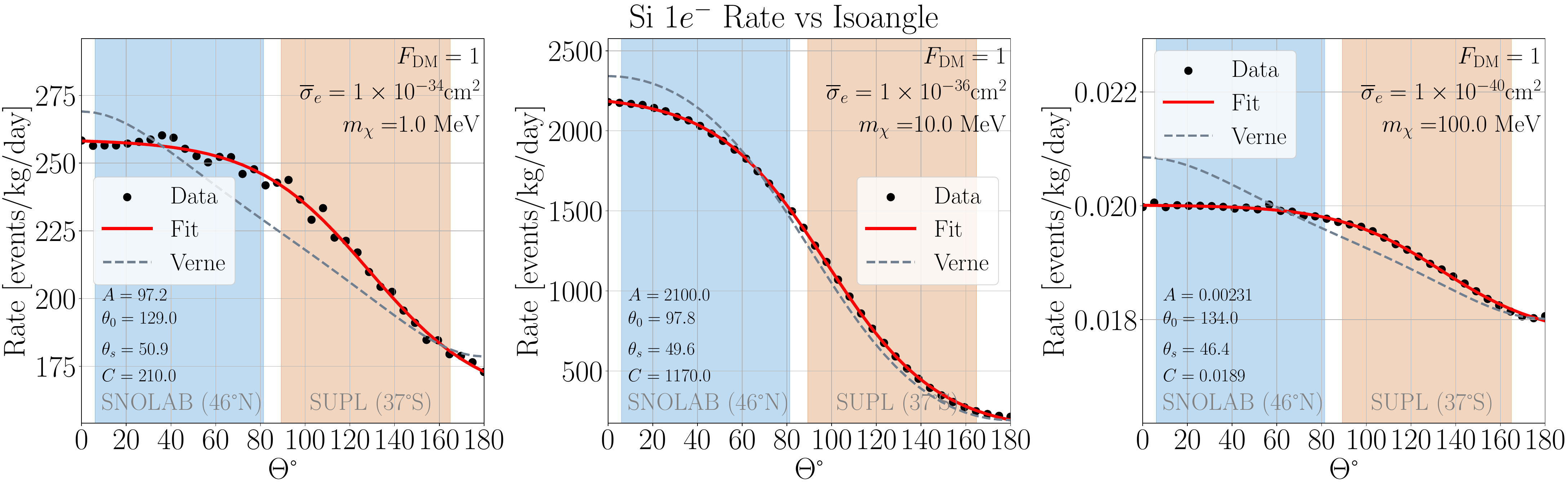}%

  \includegraphics[clip,width=0.88\columnwidth]{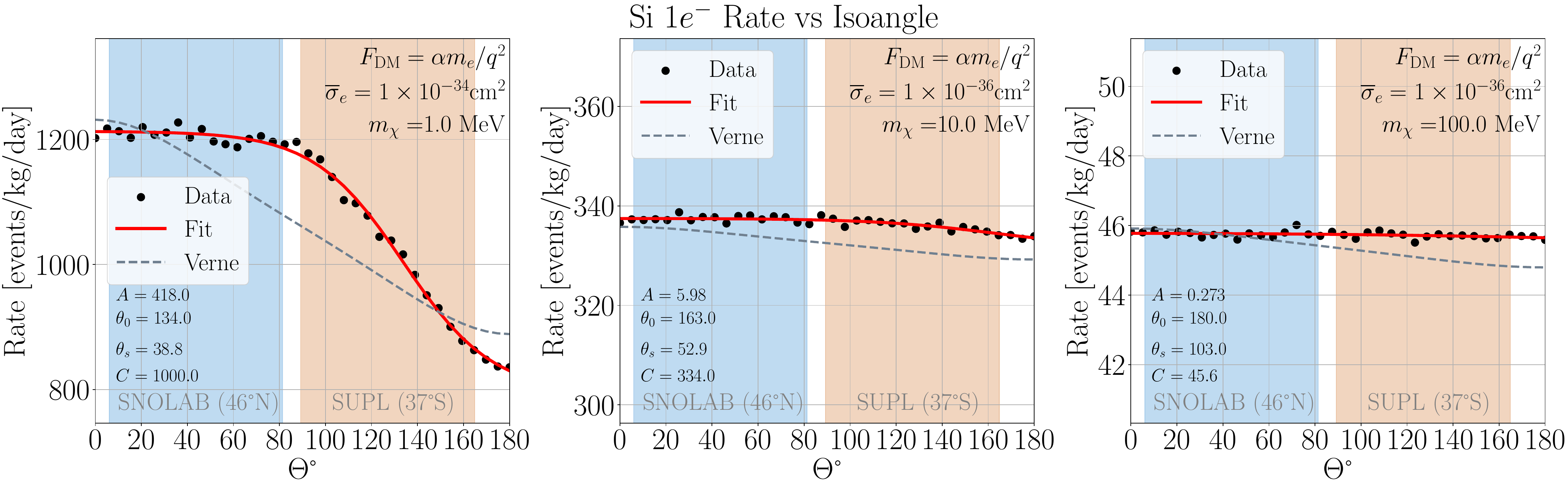}%

\caption{Scattering rate for $m_\chi=1$ MeV ({\bf left}), 10 MeV ({\bf middle}), and 100 MeV ({\bf right}) in silicon for the $n_e=1$ bin for the indicated scattering cross-sections, which are near the current constraints. 
The {\bf {\color{SNOLAB}blue}} and {\bf {\color{SUPL}orange}} regions indicate the isoangles spanned by our two representative locations, SNOLAB and SUPL, respectively. Here the data are from {\tt DaMaSCUS} simulations, while the ({\bf \color{red} red line}) is the fit to these data using Eq.~\eqref{eq: hyptan fit}. Results from {\tt Verne} are shown with a {\bf{\color{grey} grey dashed line}}.}
\label{fig:silicon_rates}

\end{figure*}

\begin{figure*}[!htb]

    \subfloat[Xenon Heavy Mediator]{%
      \includegraphics[clip,width=0.95\columnwidth]{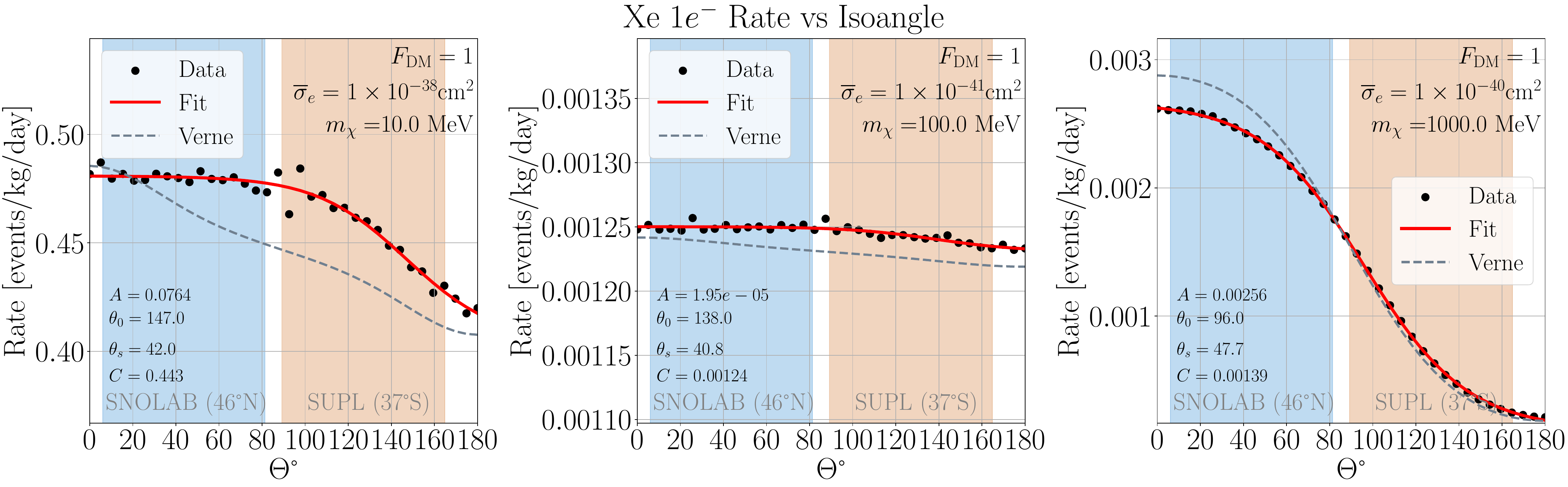}%
    }
\vskip 1mm

    \subfloat[Xenon Light Mediator]{%
      \includegraphics[clip,width=0.95\columnwidth]{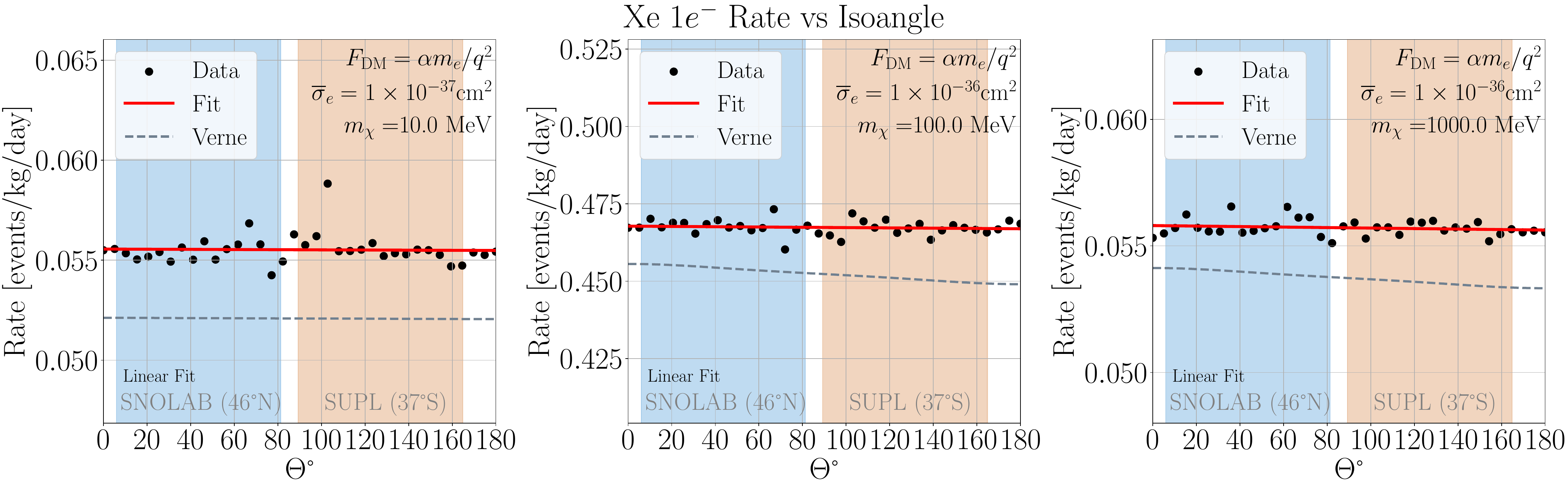}%
    }
\vskip 1mm
    
    \subfloat[Argon Heavy Mediator]{%
      \includegraphics[clip,width=0.95\columnwidth]{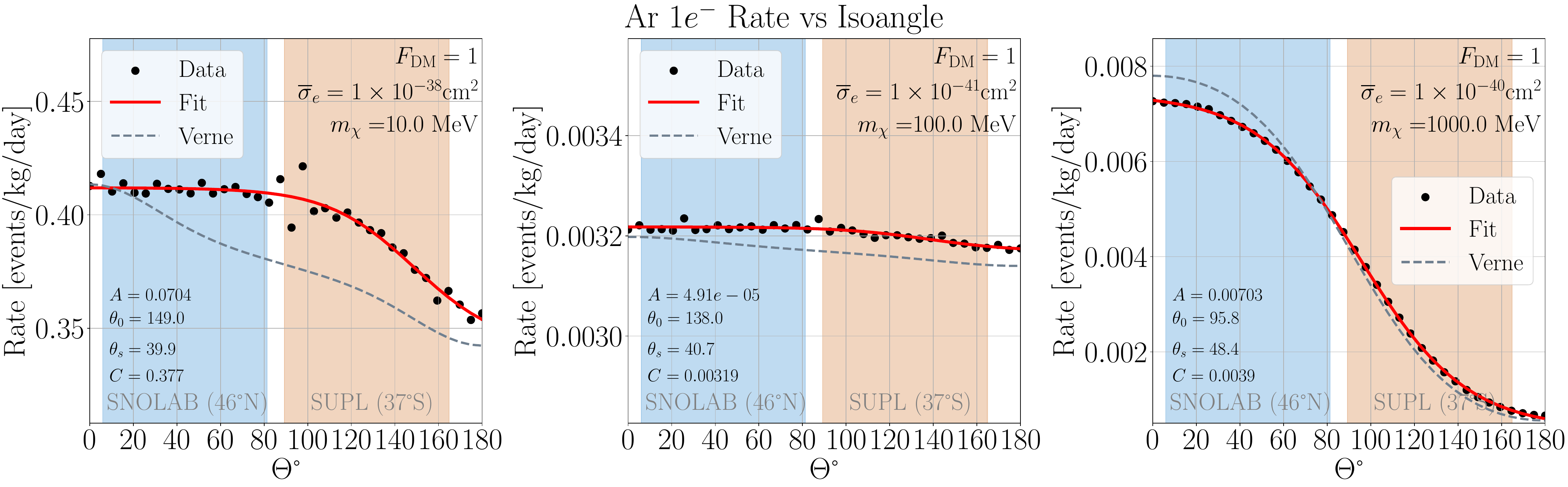}%
    }
    \vskip 1mm
    \subfloat[Argon Light Mediator]{%
      \includegraphics[clip,width=0.95\columnwidth]{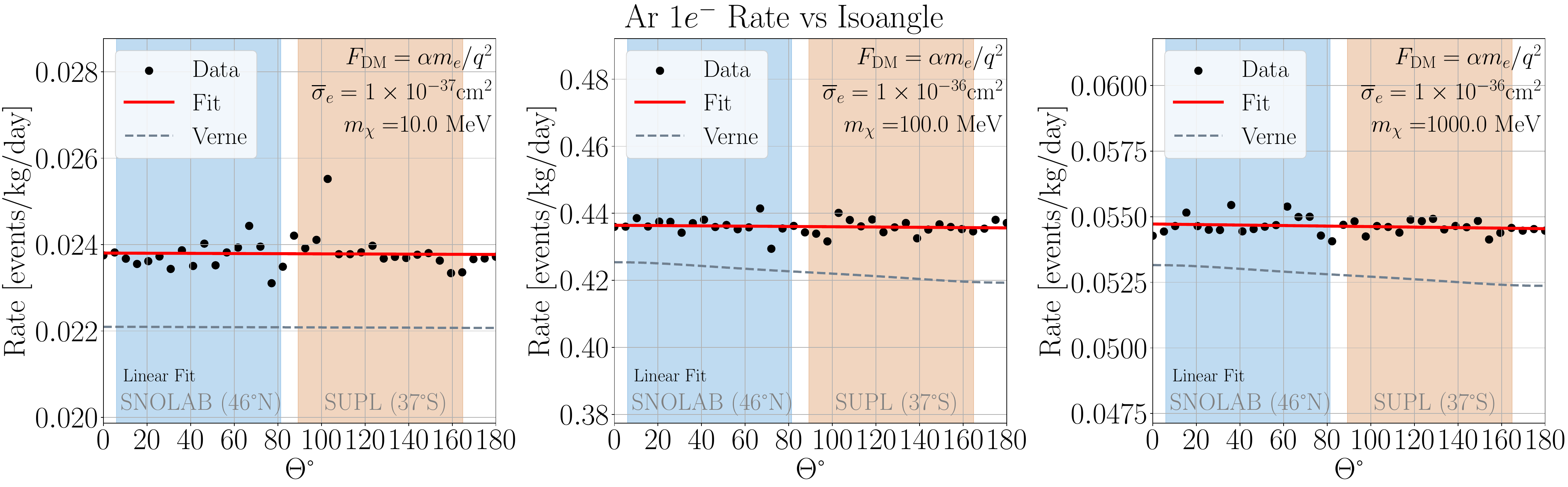}%
    }

    \caption{Same as Fig.~\ref{fig:silicon_rates} but for $m_\chi=10$ MeV ({\bf left}), 100 MeV ({\bf middle}), and 1000 MeV ({\bf right}) in xenon ({\bf top two rows}) and argon ({\bf bottom two rows}) for the $n_e=1$\NE~bin for indicated scattering cross-sections, which are near the current constraints.}

    \label{fig:fitted_noble_rates}
\end{figure*}

Representative results from the {\tt DaMaSCUS} simulations together with the fits using Eq.~\eqref{eq: hyptan fit} are shown in Figs.~\ref{fig:silicon_rates} and \ref{fig:fitted_noble_rates}, where we also compare these to the results obtained using {\tt Verne}. We see that {\tt Verne} and {\tt DaMaSCUS} predictions fall within 10\% of each other for all tested points. 
For {\tt DaMaSCUS}, the statistical uncertainty in the data points increases as the mean free path decreases, due to the increased number of samples needed to get the shape to converge. As a result, for points where the modulation is weak (see the light mediator in Fig.~\ref{fig:fitted_noble_rates}) we see that {\tt DaMaSCUS} results are dominated by statistical uncertainty, whereas with {\tt Verne} it is possible to study small modulation signals. 

Additionally we point out that {\tt Verne}, due to its analytic approximation, always predicts an inflection point at $\Theta_0 = 90^\circ$, whereas results from {\tt DaMaSCUS} suggests $\Theta_0$ may differ from $90^\circ$ depending on the DM mass and cross-section. This can impact the prediction for the expected sensitivity of detectors at different locations on Earth. For example, {\tt DaMaSCUS} would predict $\sim0$ modulation for $m_\chi=1$ MeV, $\overline\sigma_e=10^{-34}$ cm$^2$ for a detector at SNOLAB and a very strong modulation at SUPL, whereas {\tt Verne} would predict a similar modulation amplitude for both locations (see {\it e.g.} Fig.~\ref{fig:silicon_rates}).

For simplicity, the rate calculation is done over one sidereal day to obtain the correct average exposure to each direction of the DM wind.  The choice of velocity of the Earth around the Sun is important, as it affects the total scattering rate as well as the precise DM mass threshold. For our analysis, we chose the Earth's velocity on March 8th, as this is the date when the Earth's galactocentric velocity is the same as the solar galactocentric velocity~\cite{Baxter:2021pqo} (Earth's relative velocity increases to a maximum on June 8th, and a minimum on December 8th). We choose this date to average over the effect from annual modulation but study the small differences that arise from this effect in Section~\ref{subsec:significance}.

\clearpage

\subsection{Modulation Analysis}

For the modulation analysis, we define two parameters that are of interest for experiments searching for a modulation signal. The first parameter is the modulation amplitude, defined as
\begin{align}
A = \frac{R_{\rm max} - R_{\rm min}}{2}    \ ,
\end{align}
where $R_{\rm max}$ ($R_{\rm min}$) is the maximum (minimum) rate over the sidereal day time period. The amplitude $A$ provides information on the number of additional events above the average  expected for a fixed exposure at a given mass, cross-section, and location on Earth. 
The second parameter is the fractional modulation amplitude defined as
\begin{align}
f_{\rm mod} = \frac{R_{\rm max} - R_{\rm min}}{2  R_{0}}    \ ,
\end{align}
where $R_0$ is the average rate over the sidereal day. 
A significant difference above and below the constant background rate threshold must be observed in order to discover DM through modulation, even if the rates themselves are non-zero. Both $A$ and $f_{\rm mod}$ depend on geospatial coordinates, as can be inferred from Figs.~\ref{fig:silicon_rates} and~\ref{fig:fitted_noble_rates}.

\begin{figure*}[!htb]
    \includegraphics[width=\textwidth]{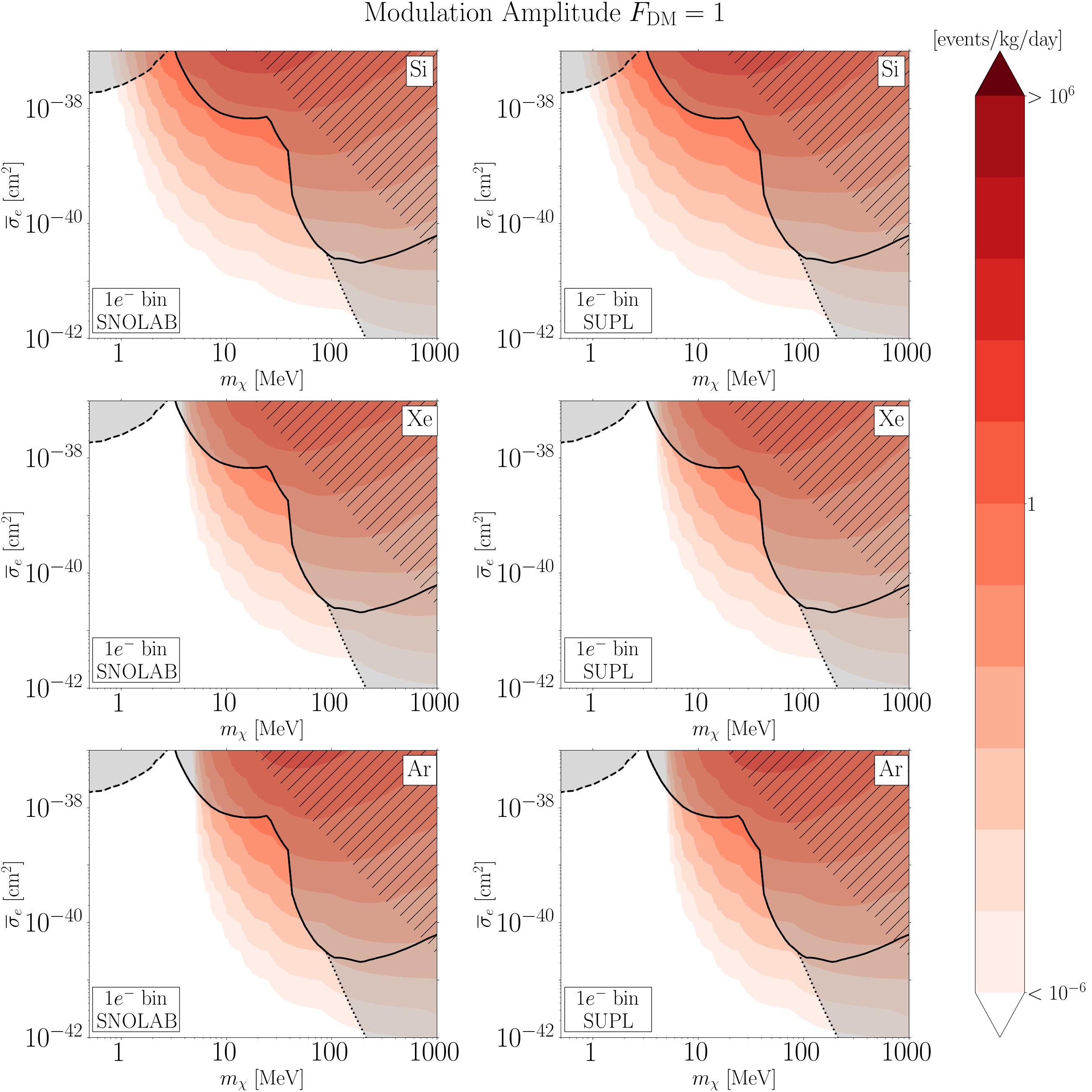}
    \caption{Modulation Amplitude [events/kg/day] from {\tt Verne} for $F_{\rm DM}=1$ (a heavy dark-photon mediator) in silicon ({\bf row 1}), xenon ({\bf row 2}), and argon ({\bf row 3}). The ({\bf left}) column is for a detector located at SNOLAB, Canada, and the ({\bf right}) column is for a detector located at SUPL in Stawell, Australia. The {\bf black, solid} curve denotes halo constraints~\cite{damicmcollaboration2025probingbenchmarkmodelshiddensector,DamicModArnquist_2024,SENSEI:2024yyt,PandaXTLi_2023,XENON:2024znc,DarkSide:2022knj}. The {\bf black, dotted} curve shows constraints from the Migdal effect from~\cite{PandaXTLi_2023}. The {\bf black, dashed} curve shows the solar reflected DM constraints from~\cite{PhysRevLett.123.251801,Emken:2024nox,XENON:2024znc}. The {\bf hashed} region is where the mean free path of the DM through the Earth is less than $R_\oplus$, where the approximations used in {\tt Verne} break down.}

    \label{fig:Modulation Amplitude FDM1}
\end{figure*}

\begin{figure*}[!htb]
    \includegraphics[width=\textwidth]{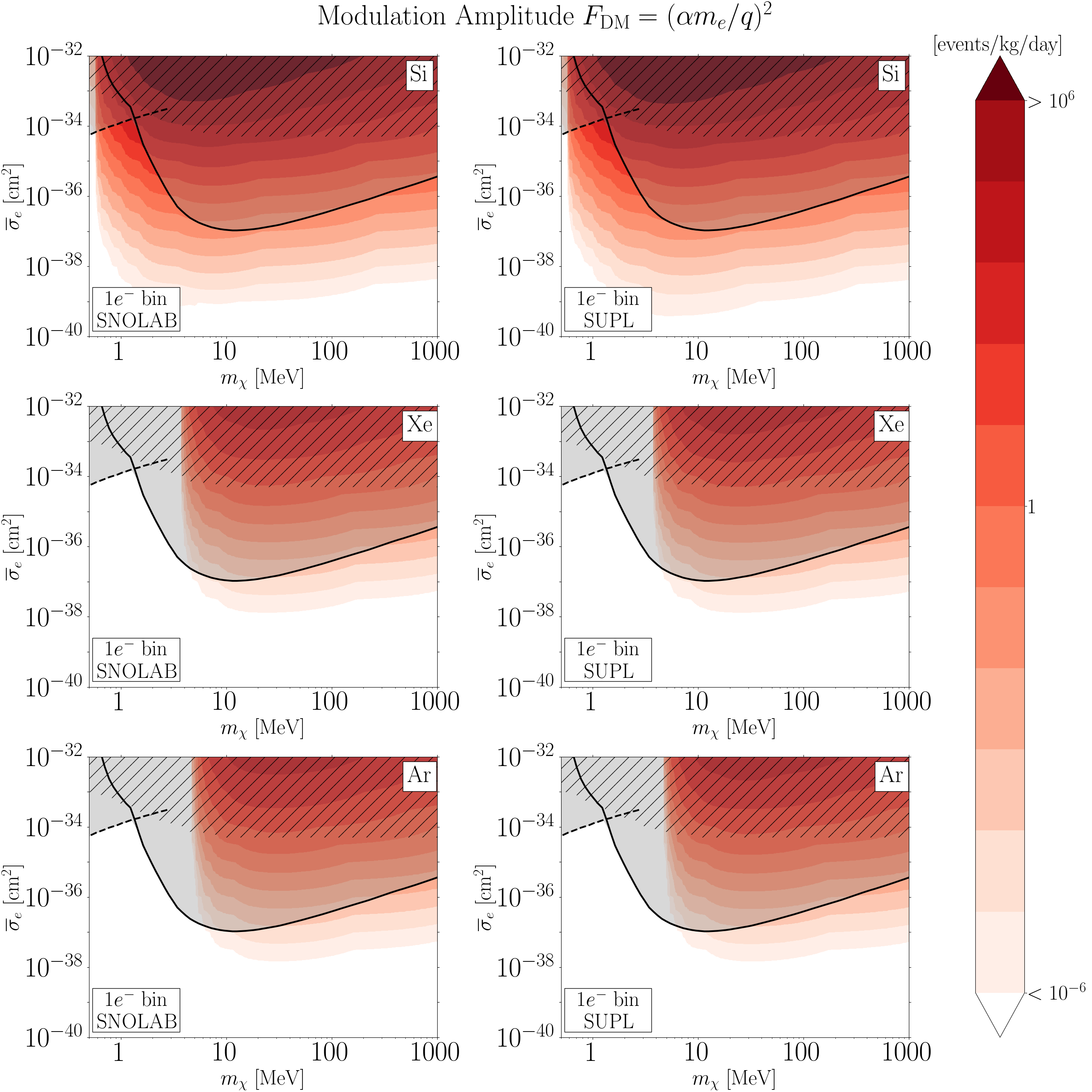}
    \caption{
    Same as Fig.~\ref{fig:Modulation Amplitude FDM1} but for $F_{\rm DM}=(\alpha m_e/q)^2$ (a light dark-photon mediator). }
    \label{fig:Modulation Amplitude FDMq2}
\end{figure*}

\begin{figure*}[!htb]
    \includegraphics[width=\textwidth]{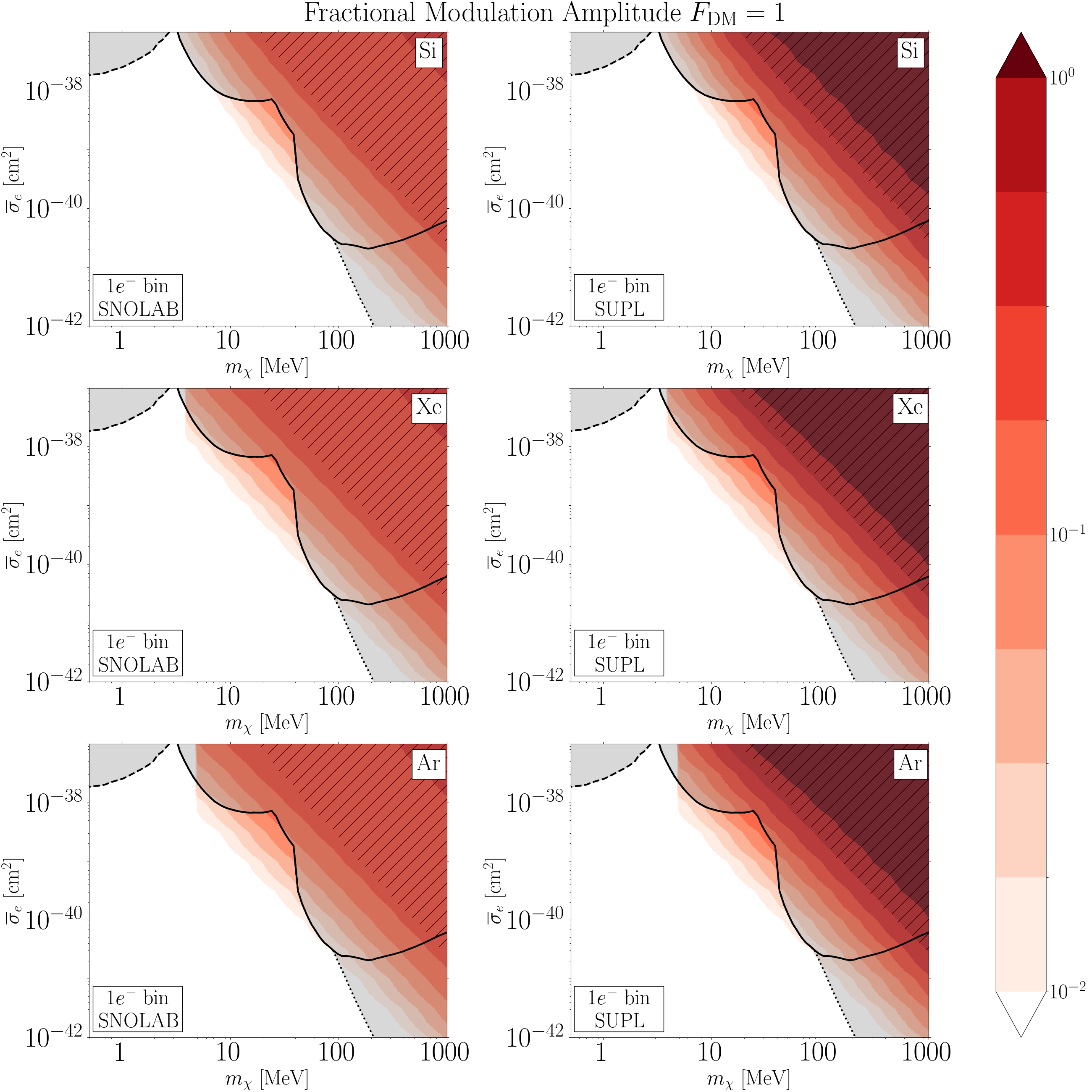}
    \caption{Fractional Modulation Amplitude from {\tt Verne} for $F_{\rm DM}=1$ (a heavy dark-photon mediator) in silicon ({\bf row~1}), xenon ({\bf row~2}), and argon ({\bf row~3}). The ({\bf left}) column is for a detector located at SNOLAB, Canada, and the ({\bf right}) column is for a detector located at SUPL in Stawell, Australia. The {\bf black, solid} curve denotes the halo constraints~\cite{damicmcollaboration2025probingbenchmarkmodelshiddensector,DamicModArnquist_2024,SENSEI:2024yyt,PandaXTLi_2023,XENON:2024znc,DarkSide:2022knj}. The {\bf black, dotted} curve shows constraints from the Migdal effect from~\cite{PandaXTLi_2023}. The {\bf black, dashed} curve shows the solar reflected DM constraints from~\cite{PhysRevLett.123.251801,Emken:2024nox,XENON:2024znc}. The {\bf hashed} region is where the mean free path of the DM through the Earth is less than $R_\oplus$, where the approximations used in {\tt Verne} break down.}
    \label{fig:Fractional Modulation Amplitude FDM1}
\end{figure*}

\begin{figure*}[!htb]
    \includegraphics[width=\textwidth]{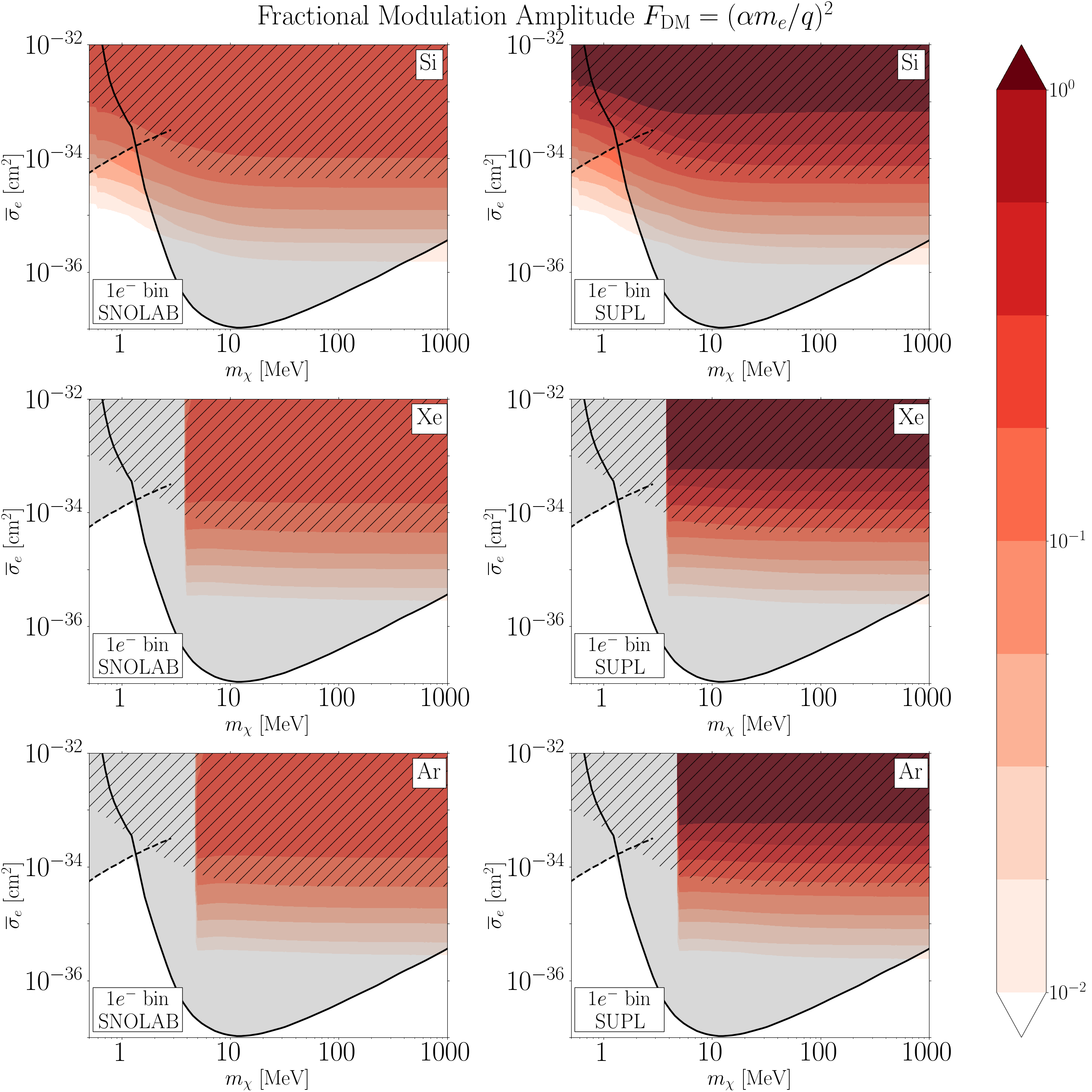}
    \caption{Same as Fig.~\ref{fig:Fractional Modulation Amplitude FDM1} but for $F_{\rm DM}=(\alpha m_e/q)^2$ (a light dark-photon mediator). }
    \label{fig:Fractional Modulation Amplitude FDMq2}
\end{figure*}

We show the modulation amplitude contours for the $1e^-$ bin in Figs.~\ref{fig:Modulation Amplitude FDM1} and \ref{fig:Modulation Amplitude FDMq2}, and the fractional modulation amplitude contours in Figs.~\ref{fig:Fractional Modulation Amplitude FDM1} and \ref{fig:Fractional Modulation Amplitude FDMq2}, for silicon, xenon, and argon, and for our two representative locations (SNOLAB, Canada, and SUPL, Australia).
These figures are calculated using {\tt Verne} for computational efficiency. We have marked the region where the DM mean free path is smaller than the radius of the Earth with hashes; here, the approximation used by {\tt Verne} breaks down. However, we note that near the current constraints, the {\tt Verne} approximation is expected to be accurate for nearly all masses and cross-sections. 

Overall, the modulation amplitudes shown in Figs.~\ref{fig:Modulation Amplitude FDM1} and~\ref{fig:Modulation Amplitude FDMq2}, as well as the fractional modulation amplitudes in Figs.~\ref{fig:Fractional Modulation Amplitude FDM1} and~\ref{fig:Fractional Modulation Amplitude FDMq2}, appear to be comparable between SNOLAB and SUPL in the regions near current constraints. As expected, we see in general the contour lines follow the same shape as Fig.~\ref{fig:mfp}, where both the modulation and fractional amplitudes flatten out for larger masses in the light mediator case but continue to increase as the DM mass gets larger in the heavy mediator case. Interestingly, we see that for the DM masses near the threshold, the modulation amplitude is relatively high for cross-section values near the current constraints, so that experiments are able to improve upon current bounds by searching for modulation. 

\newpage

\newpage

\newpage

\clearpage

\begin{figure*}[!htb]
    \includegraphics[width=0.49\textwidth]{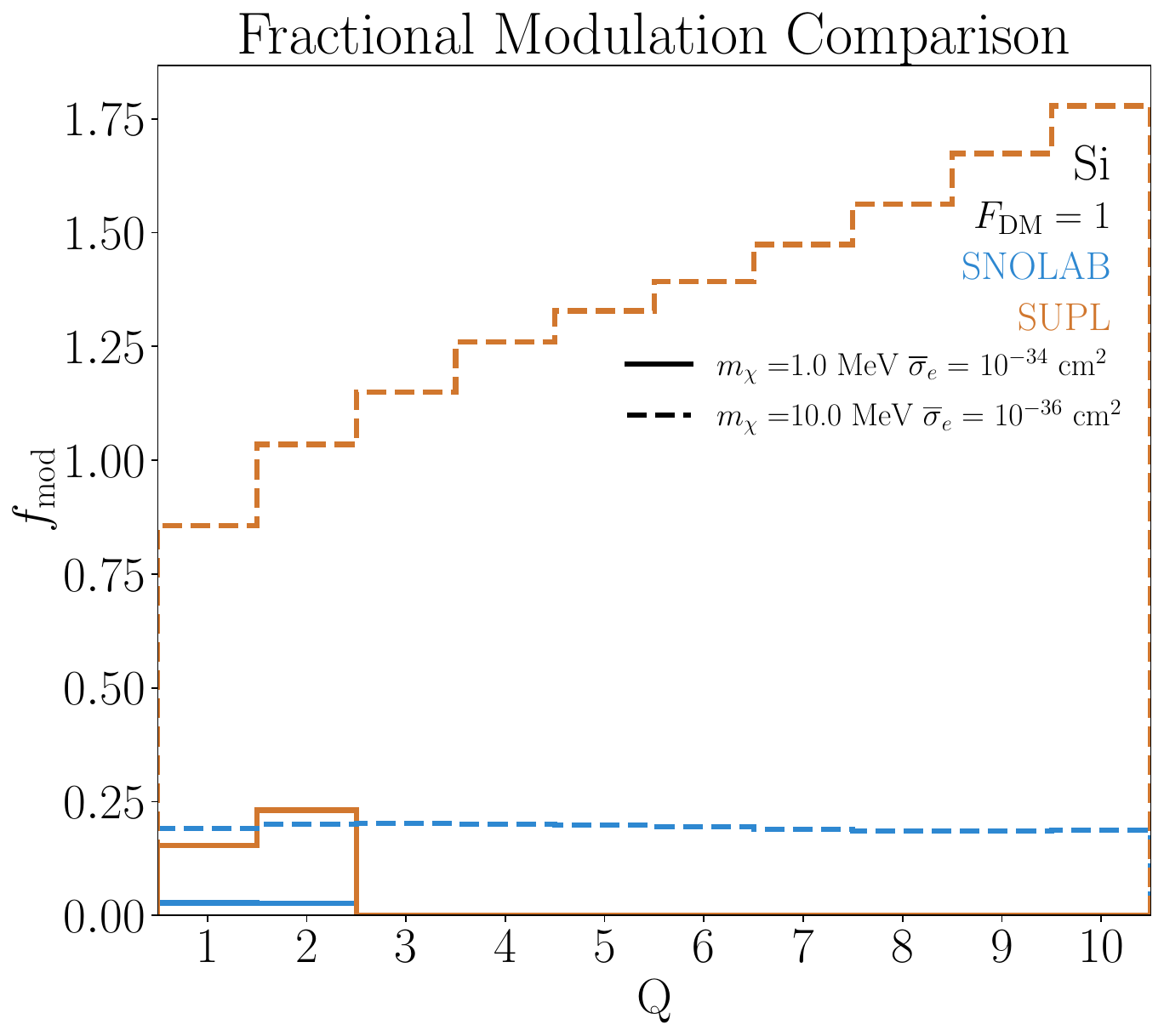} \includegraphics[width=0.49\textwidth]{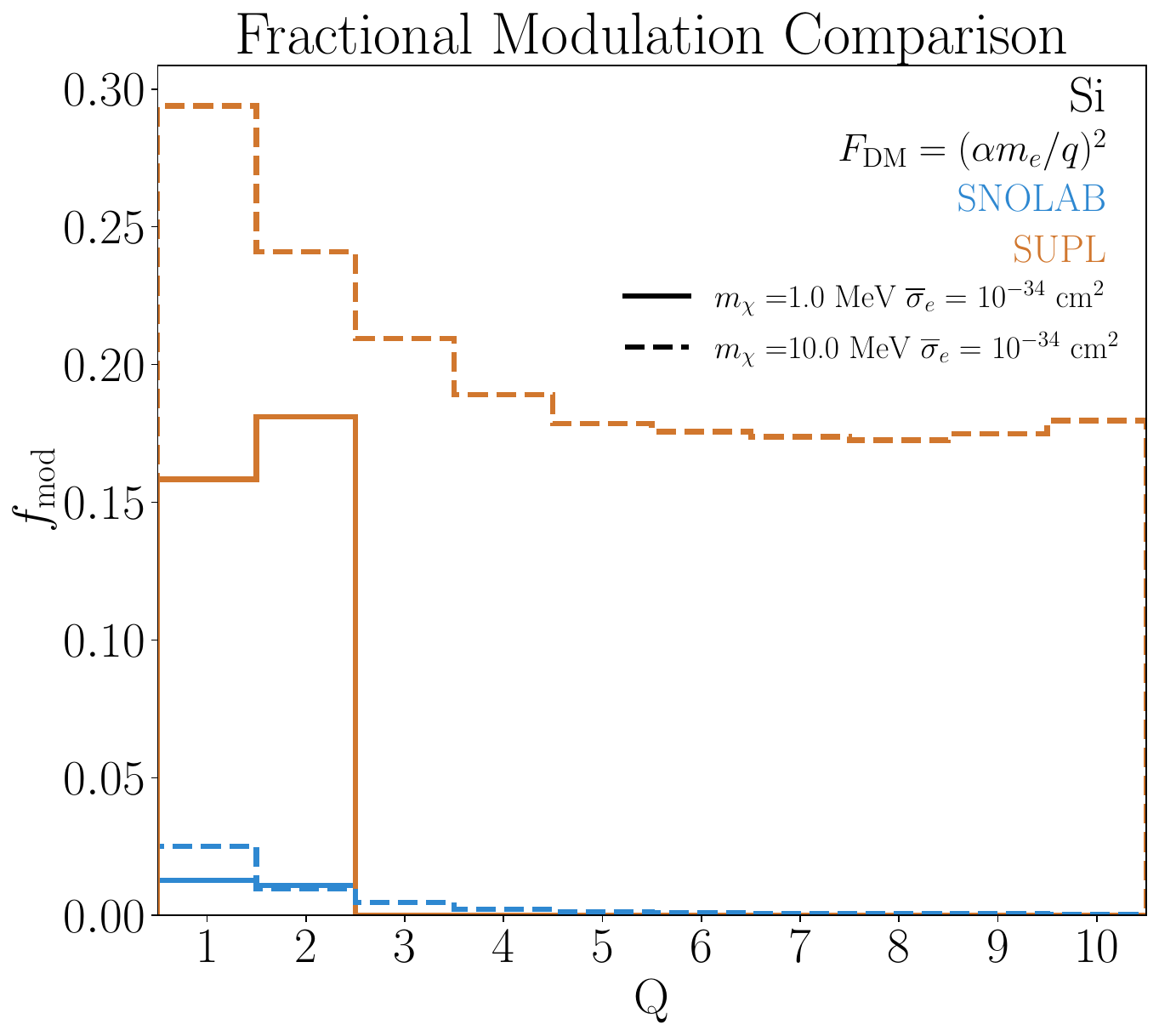}
    \caption{Modulation amplitude $f_{mod}$ versus number of electrons $Q$ in silicon for DM interacting with electrons through a  heavy ({\bf left}) or light ({\bf right}) dark photon mediator, using results from {\tt DaMaSCUS} for 1~MeV ({\bf solid}) and 10~MeV ({\bf dashed}) for a detector located at SNOLAB ({\bf{\color{SNOLAB} blue}}) and SUPL ({\bf\color{SUPL} orange}). The cutoff above $Q = 2$ for $m_\chi=1$ MeV is due to kinematics. One can clearly see that the fractional modulation is dramatically stronger at SUPL, but overall is not strongly dependent on Q, especially in the light-mediator case. This shows that modulation is fairly consistent across all bins.
    }
    \label{fig:silicon_frac_nebins}
\end{figure*}

\begin{figure*}[!t]
    \includegraphics[width=0.49\textwidth]{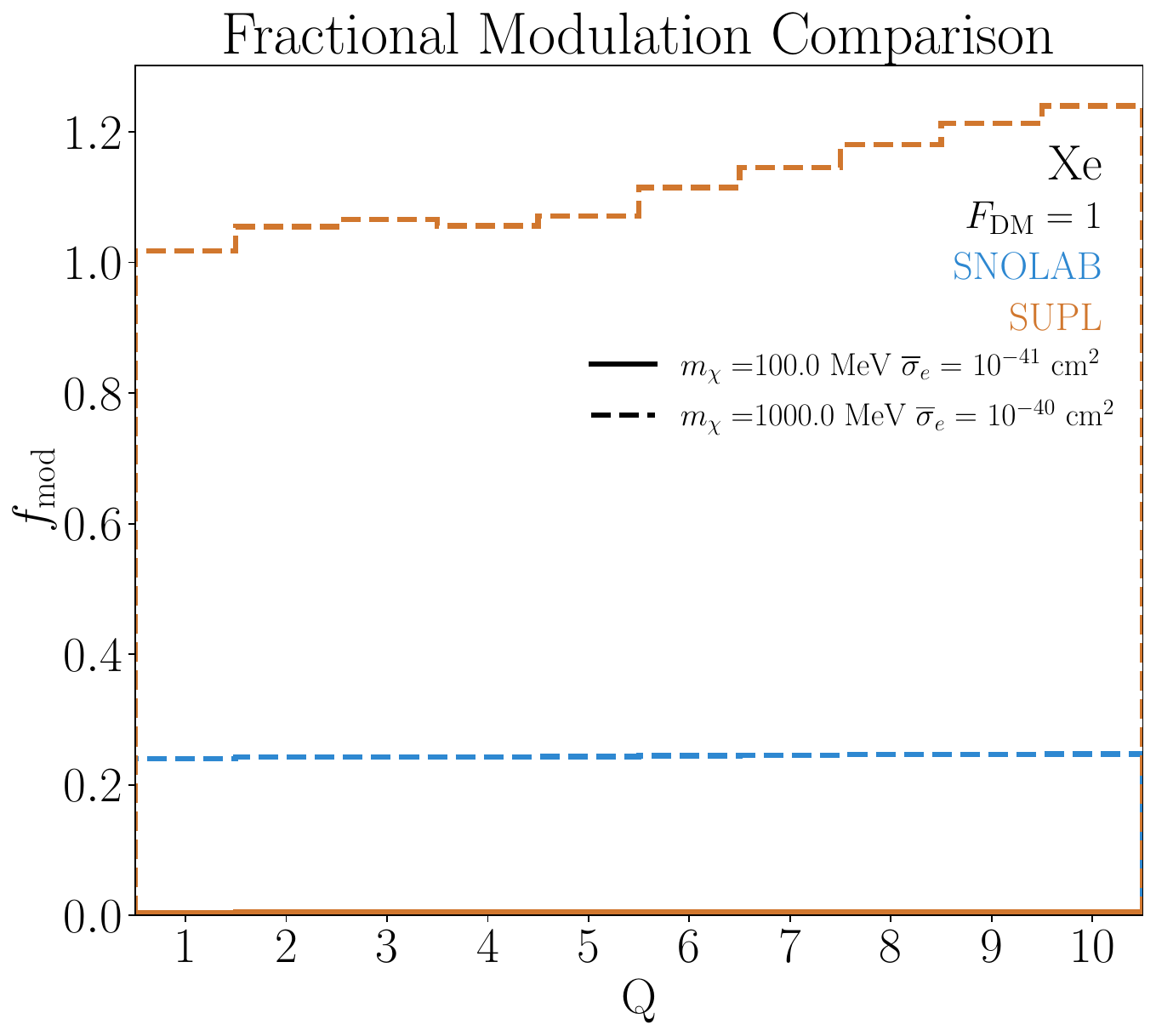} \includegraphics[width=0.49\textwidth]{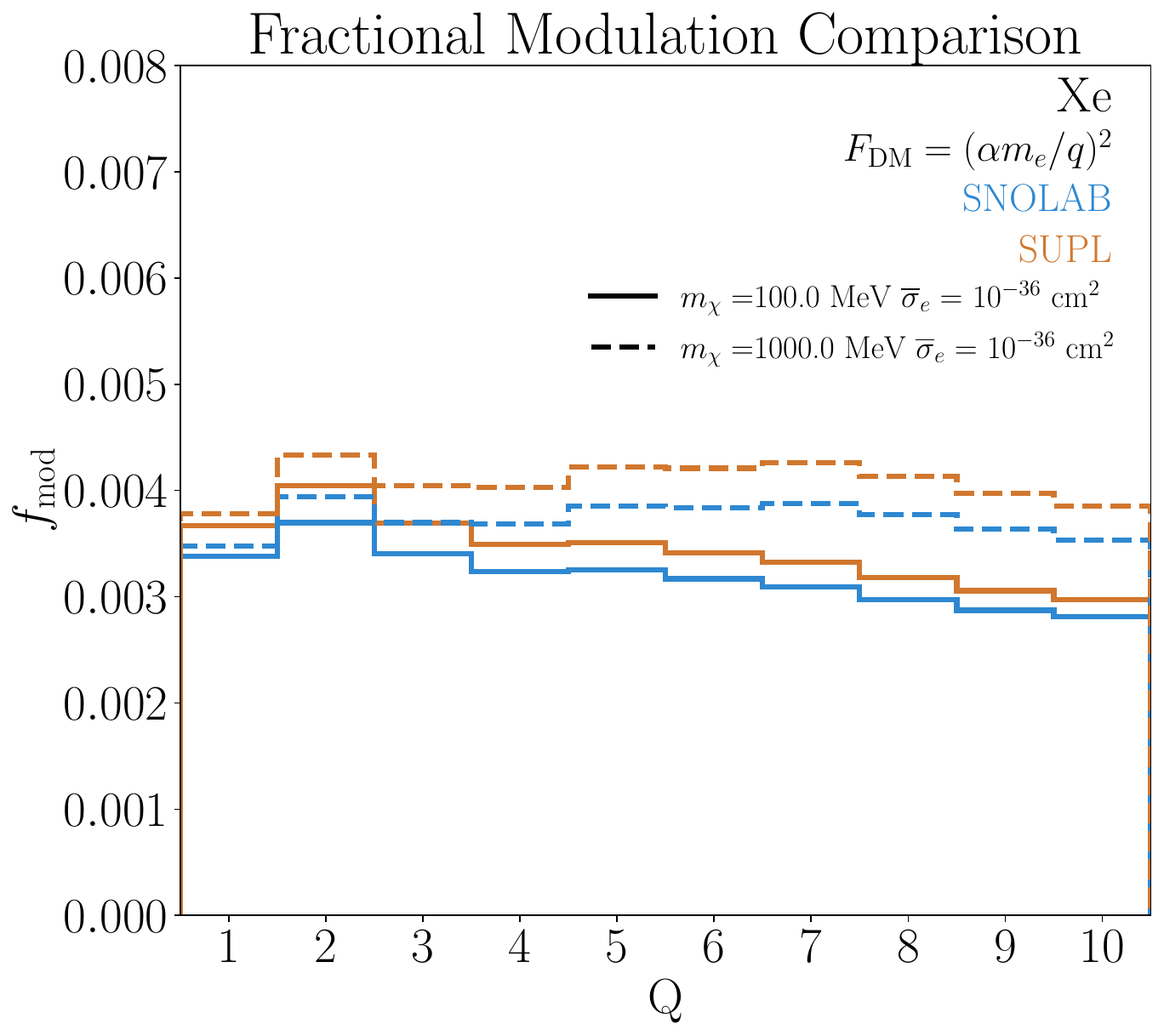}
    \caption{Modulation amplitude $f_{mod}$ versus number of electrons $Q$ in xenon for DM interacting with electrons through a heavy ({\bf left}) or light ({\bf right}) dark photon mediator using results from {\tt Verne}, which was used instead of {\tt DaMaSCUS} due to simulation limitations at larger masses. Shown are two masses, 100~MeV ({\bf solid}) and 1000~MeV ({\bf dashed}), for both SNOLAB ({\bf{\color{SNOLAB} blue}}) and SUPL ({\bf \color{SUPL} orange}). One can clearly see that the fractional modulation is stronger at SUPL, but overall is not strongly dependent on Q, especially in the light mediator case.
    }
    \label{fig:xenon_frac_nebins}
\end{figure*}

\begin{figure*}[!t]
    \includegraphics[width=0.49\textwidth]{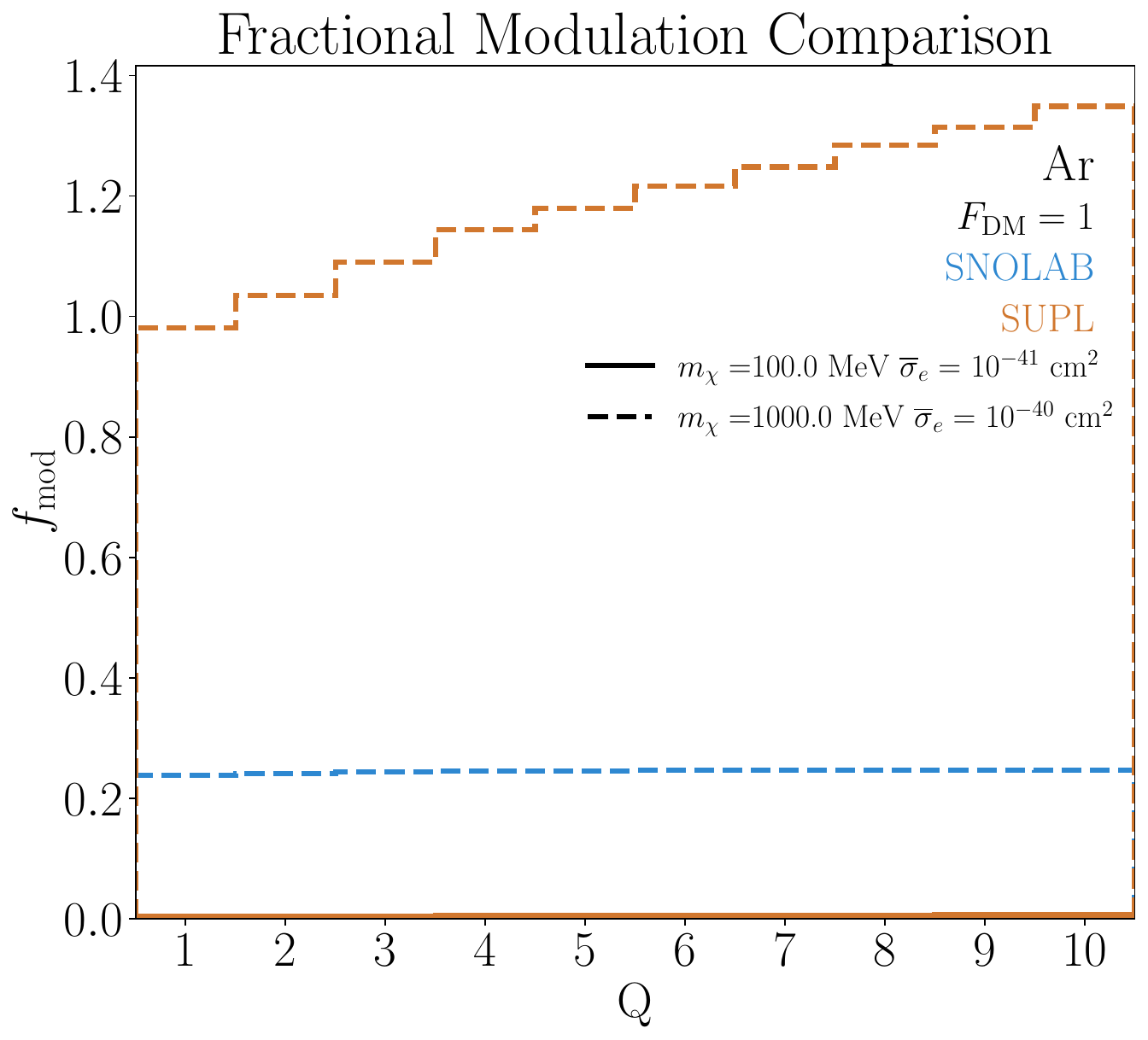} \includegraphics[width=0.49\textwidth]{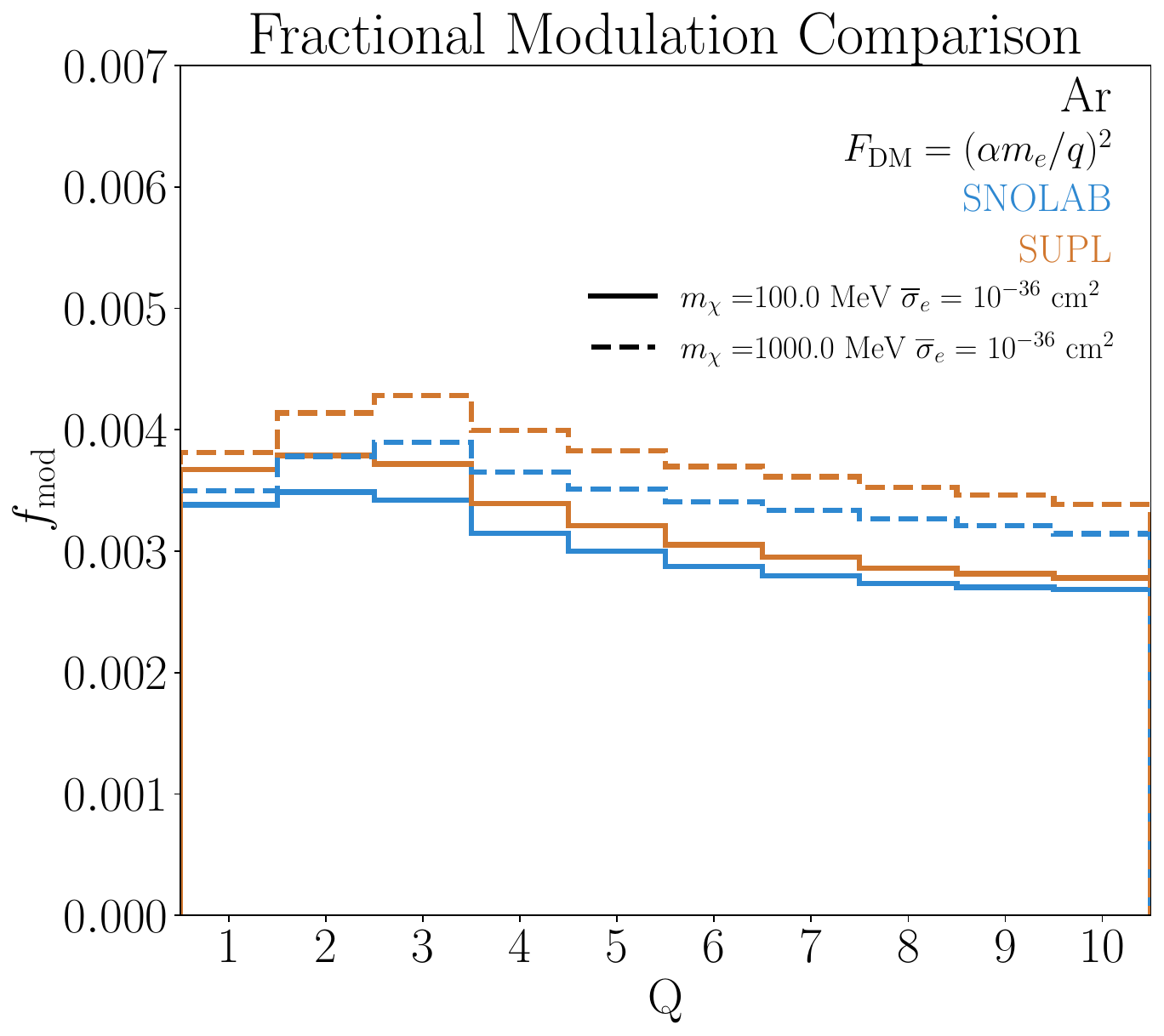}
    \caption{Same as \ref{fig:xenon_frac_nebins} but for argon.}
    \label{fig:argon_frac_nebins}
\end{figure*}

In Figs.~\ref{fig:silicon_frac_nebins},~\ref{fig:xenon_frac_nebins}, and~\ref{fig:argon_frac_nebins}, we plot the fractional amplitudes for different \NE~bins in our two representative locations for two representative masses. While we see a stark difference between the two locations, we also see that the fractional amplitude stays relatively constant for most \NE~bins for both the heavy and light mediators. This suggests that one can search for this modulation in any bin that provides a large enough signal and has a constant background, as the fractional modulation is not particularly sensitive to the choice of bin. This is markedly different from the annual fractional modulation amplitude spectrum shown in~\cite{Essig:2015cda}.

While Figs.~\ref{fig:silicon_rates},~\ref{fig:fitted_noble_rates},~\ref{fig:silicon_frac_nebins},~\ref{fig:xenon_frac_nebins},~\ref{fig:argon_frac_nebins} indicate some location dependence of the rate amplitudes, this effect is more pronounced (see Figs.~\ref{fig:Fractional Modulation Amplitude FDM1},~\ref{fig:Fractional Modulation Amplitude FDMq2}) in the high-mass and/or large cross-section regimes, which are already excluded. As one can see in Figs.~\ref{fig:Modulation Amplitude FDM1},~\ref{fig:Modulation Amplitude FDMq2} the amplitude is very similar in both two locations in parameter space that is currently unconstrained. Therefore, we conclude that a modulation analysis remains viable in either hemisphere.

For silicon detectors, much of the accessible parameter space for a heavy mediator has been excluded. However, in the case of a light mediator, we observe relatively large modulation amplitudes for masses $m_\chi<10$ MeV near the current experimental limits. This suggests that ongoing experiments have the sensitivity to extend these constraints, as demonstrated in~\cite{DamicModArnquist_2024}. Notably, this lies within the 1\NE~regime, which has so far been dominated by backgrounds. Until these backgrounds are better characterized, modulation searches offer a promising alternative for probing these low DM masses.

The noble gas detectors have an advantage over the silicon detectors due to their significantly larger target mass. Since these experiments have several bins with large amounts of backgrounds, we see that they could potentially improve their sensitivity to DM-electron scattering across a wider energy spectrum.

\clearpage
\subsection{Significance Reach in Silicon}\label{subsec:significance}
Identifying a signal requires sufficient exposure to determine whether events in a given bin arise from signal plus background or from background alone. A useful quantity is the significance $\sigma$ defined as 
\begin{align}
\sigma = \frac{\Delta S}{\sqrt{S_{tot} + B}}    \ ,
\end{align}
where $\Delta S \equiv f_{mod} S_{tot}$ is the modulation amplitude, $S_{tot}$ is the number of signal events, and $B$ is the number of background events.  We note that the significance scales as $1/\sqrt{\mathrm{exposure}}$.

\begin{figure*}[t]
    \centering
    \includegraphics[width=0.6\textwidth]{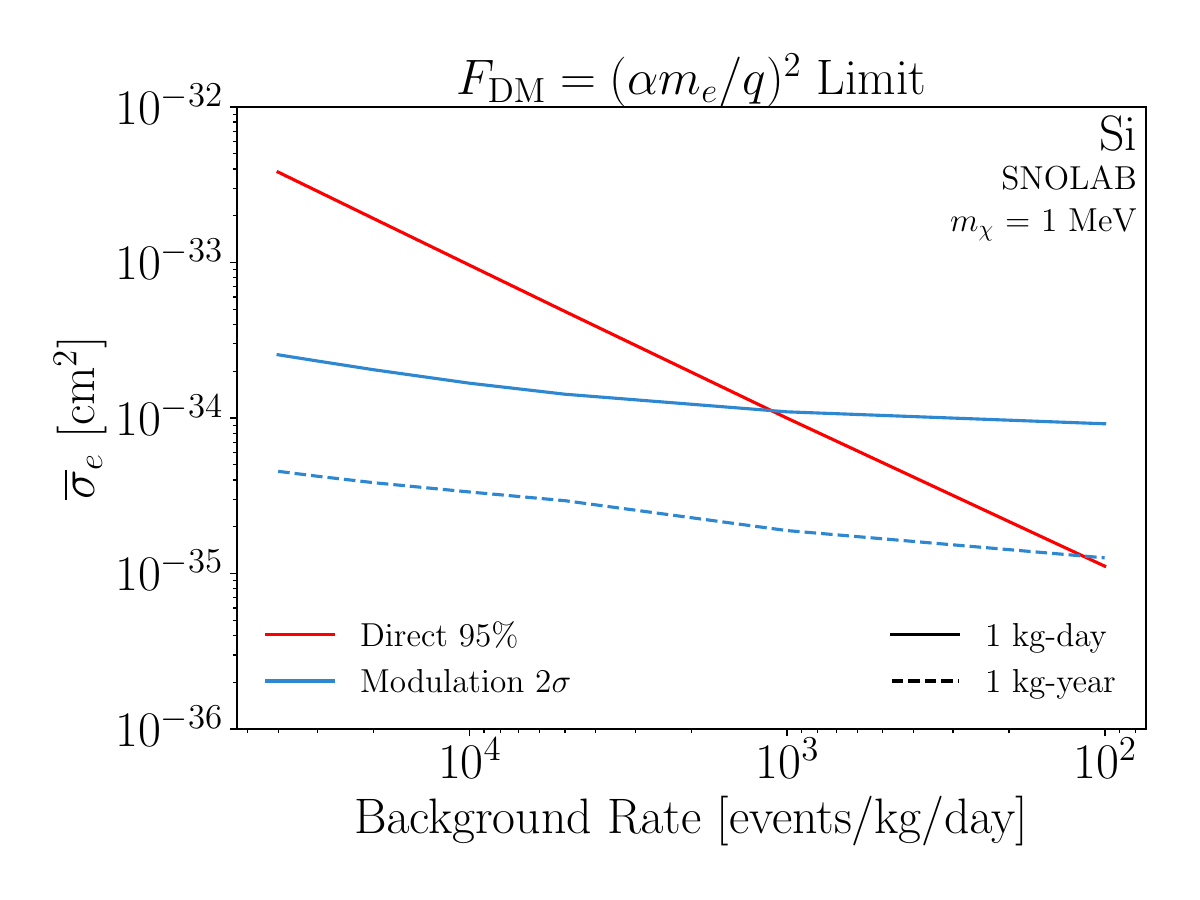}
    \caption{The 2$\sigma$ modulation discovery reach for a silicon detector located at SNOLAB for a 1~kg-day of exposure ({\bf{\color{SNOLAB} solid blue line}}) and 1~kg-year of exposure (\textbf{\color{SNOLAB} dashed blue line}), together with the 95\% confidence-level limit from the total  rate (\textbf{\color{red} solid red line}) versus the observed background rate, assuming a DM mass of 1~MeV and a light dark-photon mediator. The leftmost point on the $x$-axis is the current SENSEI 1\NE\ background rate listed in Table~\ref{table:backgroundrates}.}
    \label{fig:directvsmod}
\end{figure*}

We first plot the expected significance as a function of the background event rate for $m_\chi = 1$~MeV in Fig.~\ref{fig:directvsmod}. The leftmost value is the current SENSEI 1\NE\ background rate, taken from the ``golden quadrant'' in~\cite{SENSEI:2024yyt} and scaled up to 1 kg-day. For comparison, we also plot the expected limit from a direct measurement of the rate as a function of the background rate. We see that there is a clear transition between where the background rate is sufficiently low for a direct search to perform better than a modulation search, to where the background rate is so large that a modulation search provides a better bound.  The transition values differ for different DM masses and electron bins, but for the 1\NE\ bin a modulation search will remain viable until backgrounds can be reduced further. We expect a similar result for noble liquid detectors. 

\begin{figure*}[h]
    \centering
    \includegraphics[width=0.5\textwidth]{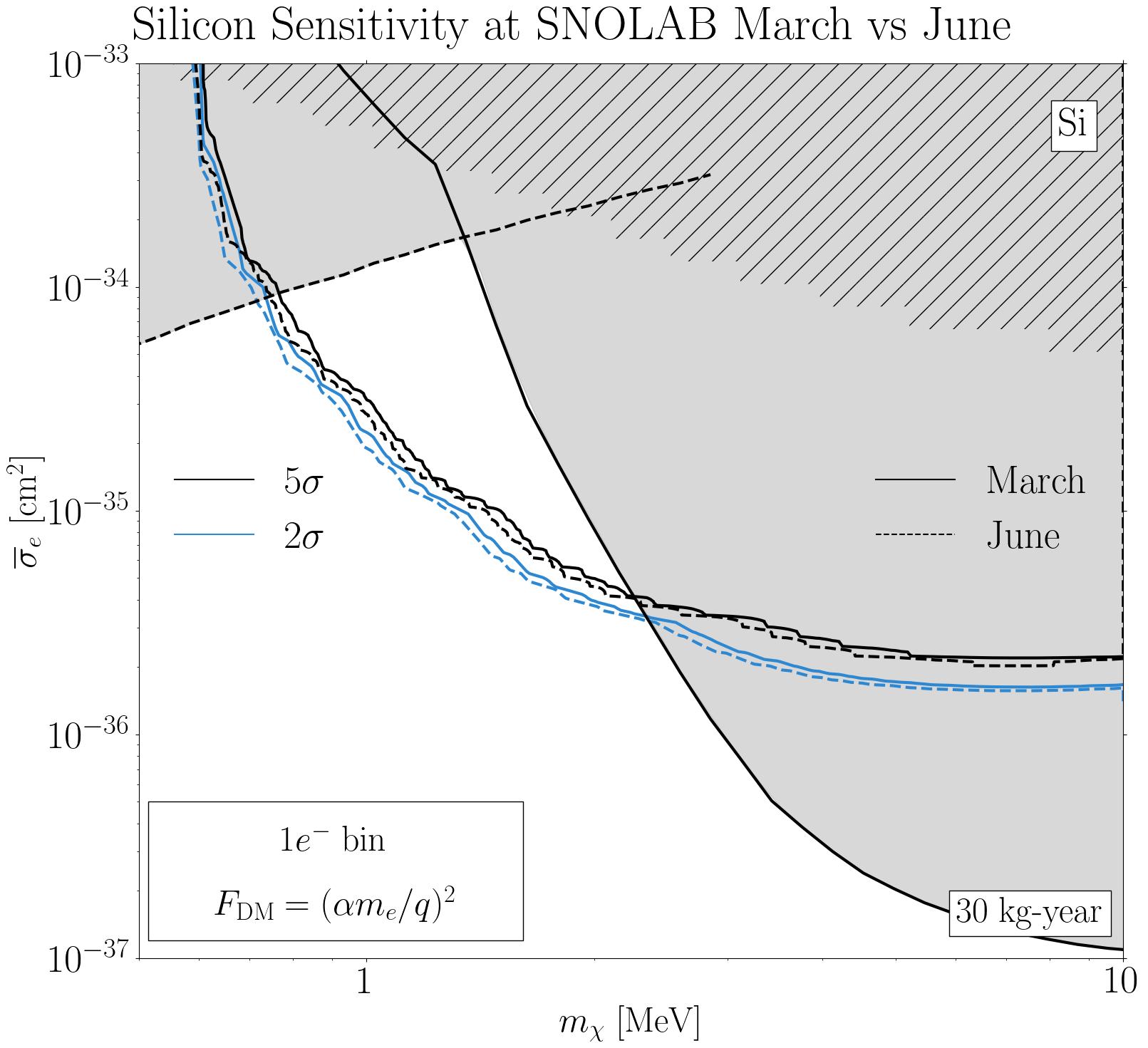}
    \caption{The 2$\sigma$ (\textbf{\color{SNOLAB} blue}) and 5$\sigma$ (\textbf{black}) modulation discovery contours of the light-mediator model for a silicon detector located at SNOLAB at a 30 kg-year exposure for March ({\bf solid}) and for June ({\bf dashed}). Halo constraints (solid black line and gray region) are combined using~\cite{damicmcollaboration2025probingbenchmarkmodelshiddensector,DamicModArnquist_2024,SENSEI:2024yyt,PandaXTLi_2023,XENON:2024znc,DarkSide:2022knj}.  Solar reflected DM constraints (black dashed line and gray region) come from~\cite{PhysRevLett.123.251801,Emken:2024nox,XENON:2024znc}. These significance contours are calculated using {\tt Verne} and the {\bf hashed} region is where the mean free path of the DM through the Earth is less than $R_\oplus$, where the approximations used in {\tt Verne} break down.}
    \label{fig:seasonal}
\end{figure*}

We also plot a comparison of the $2\sigma$ and $5\sigma$ contours in Fig.~\ref{fig:seasonal}. We note that there is not a strong difference in the discovery potential in March, where the Earth's galactocentric velocity is the same as the Sun's, vs June, where this same velocity would be at its maximum. This figure demonstrates that while there is an advantage to performing an analysis in the summer, the effect is not strong. 

\begin{figure*}[!t]
    \includegraphics[width=0.99\textwidth]{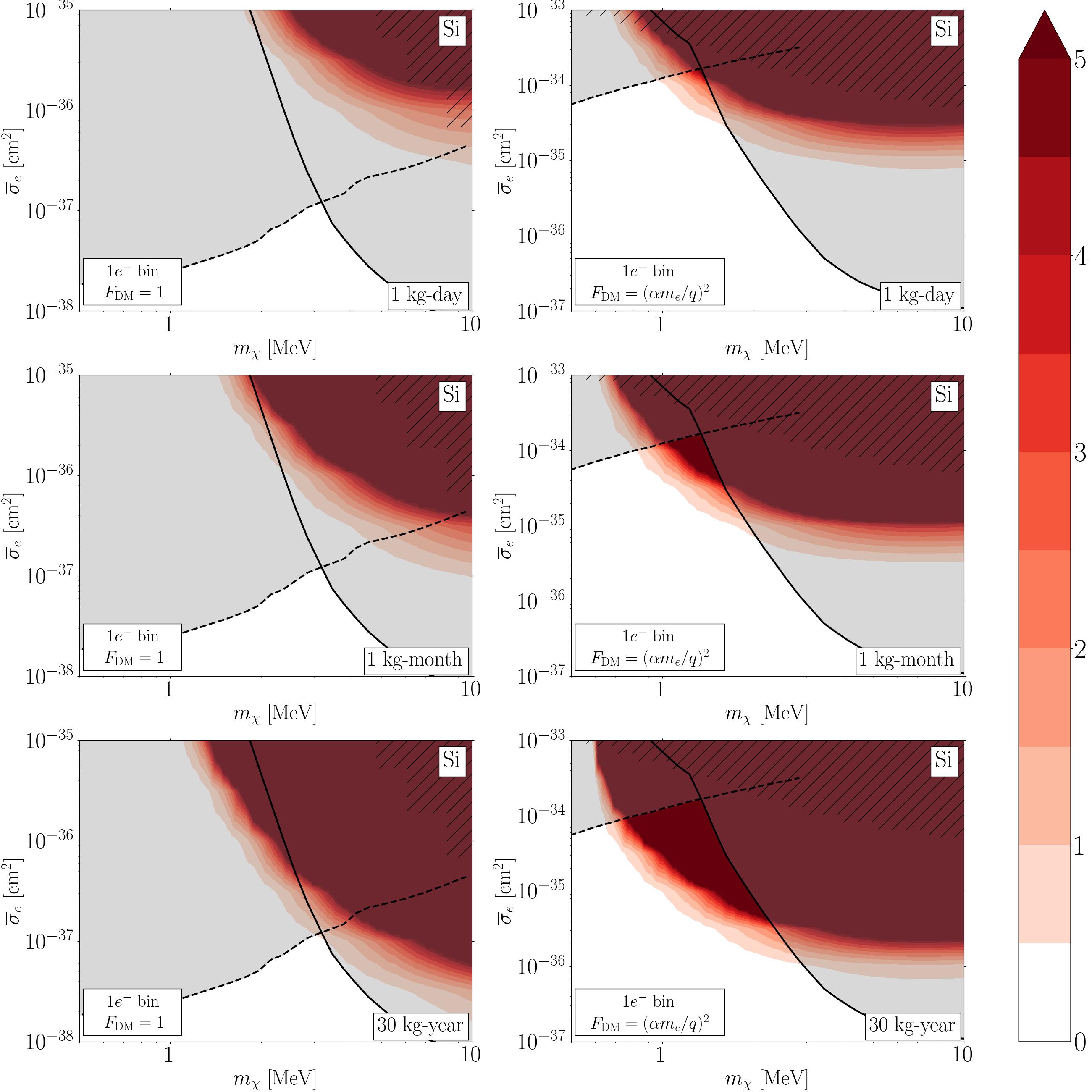}
    \caption{Expected significance contours from a modulation search in silicon for an exposure of 1 kg-day ({\bf row 1}), 1 kg-month ({\bf row 2}), and 30 kg-years ({\bf row 3}) for the 1\NE\ bin for a detector located at SNOLAB, Canada. The background 1\NE\ rate is taken from~\cite{SENSEI:2024yyt} assuming that the observed 1\NE\ rate consists entirely of background events and is listed in Table~\ref{table:backgroundrates}. The ({\bf left}) column is for a heavy dark-photon mediator, and the ({\bf right}) column is for a light dark-photon mediator. 
    Halo constraints (shaded region above solid black line) are combined using~\cite{damicmcollaboration2025probingbenchmarkmodelshiddensector,DamicModArnquist_2024,SENSEI:2024yyt,PandaXTLi_2023,XENON:2024znc,DarkSide:2022knj} combined with constraints from the Migdal effect from~\cite{PandaXTLi_2023}.  Solar reflected DM constraints (shaded region above the black dashed line) come from~\cite{PhysRevLett.123.251801,Emken:2024nox,XENON:2024znc}. These significance contours are calculated using {\tt Verne} due to simulation limitations in {\tt DaMaSCUS}.  The {\bf hashed} region is where the mean free path of the DM through the Earth is less than $R_\oplus$, where the approximations used in {\tt Verne} break down.
    }
    \label{fig:Silicon1eSignificanceSnolab}
\end{figure*}

\begin{figure*}[!htb]
    \includegraphics[width=0.99\textwidth]{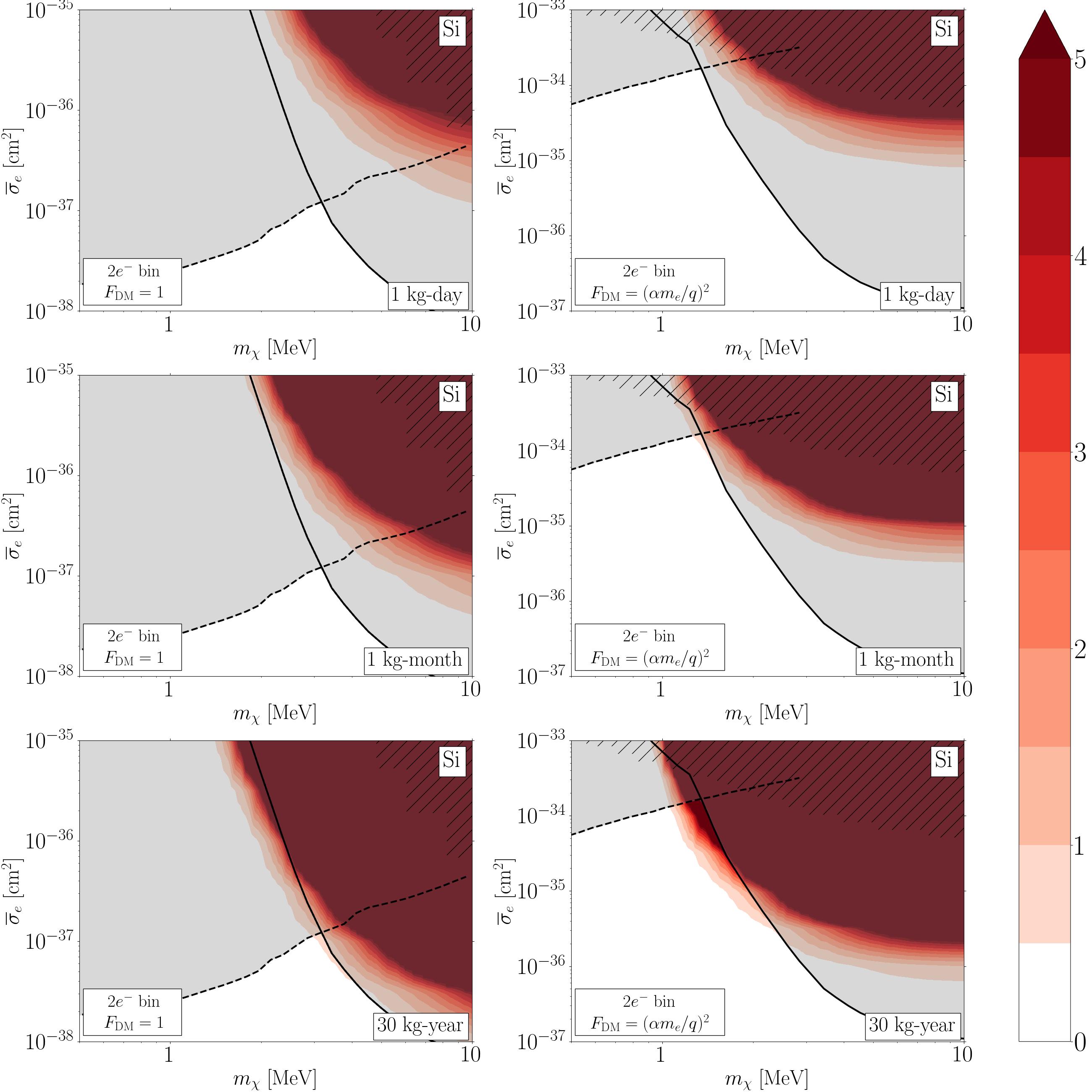}
    \caption{Same as Fig.~\ref{fig:Silicon1eSignificanceSnolab} but for the 2\NE\ bin.} 
    \label{fig:Silicon2eSignificanceSnolab}
\end{figure*}

To see where discovery reach is possible, we use the previously mentioned SENSEI 1\NE background rate from~\cite{SENSEI:2024yyt} to calculate the expected signal $\Delta S$ using {\tt QCDark} w/ screening and the results from {\tt Verne}, and plot the significance contours for the $1e^-$ bin in Fig.~\ref{fig:Silicon1eSignificanceSnolab}, for three exposures:  1~kg-day, 1~kg-month, and the target Oscura exposure of 30~kg-years~\cite{Oscura:2022vmi}. We perform a similar analysis using the $2e^-$ bin and assuming all events from~\cite{SENSEI:2023zdf} are backgrounds, and plot the results in Fig~\ref{fig:Silicon2eSignificanceSnolab}.  While much of the parameter space is already ruled out for the heavy mediator, we see that some discovery reach is possible in the low-mass regime for the light mediator, precisely in the region where current searches are background limited.
From these figures, we note that with just 1~kg-month of exposure one can detect the presence of daily modulation with $5\sigma$ in regions of parameter space that are not yet ruled out. We show results for a detector located at SNOLAB, Canada. As noted above, for probing cross-section values that are near the current bounds, the results for a detector located at SUPL, Australia are very similar to the SNOLAB results, and hence we relegate relevant figures to Appendix~\ref{app:south}.

\newpage
\newpage

\newpage

\section{Conclusion}
\label{sec:conclusion}
Low-threshold direct-detection experiments have steadily been probing the possible parameter space for low-mass DM,
but have faced challenges at the lowest DM masses due to large backgrounds.  One way to mitigate this problem is by searching for a modulating signal on top of a flat background rate. For sufficiently large DM interactions with ordinary matter, DM can scatter within the Earth, which can enhance or attenuate the flux seen at a detector. Detectors at different locations on Earth would consequently have different predictions for the expected signal rate, which will moreover modulate over the course of a sidereal day.

In order to study this daily modulation it is necessary to simulate DM scattering in the Earth.  We compare two tools, {\tt DaMaSCUS}, which provides a full 3D simulation of DM-Earth scattering, and {\tt Verne}, which is faster and makes an analytic approximation that is accurate when the mean free path of the DM through the Earth is greater than the radius of the Earth. We find that these tools provide comparable results near the current direct-detection bounds. Focusing on DM that couples to a dark-photon mediator, we use these tools to study the implications of the modulation effects for silicon, xenon, and argon targets, for a detector located at SNOLAB, Canada in the Northern hemisphere, and for a detector located at Stawell, Australia in the Southern hemisphere.  In this model, the Earth-scattering effects are dominated by DM-nucleus scattering, while DM-electron scattering dominates the signal in the experiments for the DM masses considered in our work. 

While the detector location affects the modulation amplitude, this effect is more prominent for DM masses and cross-sections that have already been disfavored by current direct-detection experiments. Near current constraints, a detector in the North and South see comparable modulation effects, meaning that no one location on Earth offers a significant advantage over another when searching for a modulation signal from Earth-scattering effects.  We also see that the fractional modulation amplitude remains approximately flat across different electron bins (we find at most a factor of two variation in silicon for a heavy dark-photon mediator between the $Q=1$\NE\ and $Q=10$\NE\ bins), meaning that experiments can use any bin that has a sufficient number of events to perform a modulation search.

Finally, we study the projected sensitivity of experiments assuming current background rates and different exposures. For silicon, we see that most of the parameter space where this search would be effective has been ruled out in the case of DM interacting with a heavy dark-photon mediator. In the case of a light dark-photon mediator, we see that experiments have discovery potential in the 1\NE\ bin, precisely where backgrounds have thus far been the most limiting. We also see that backgrounds must be reduced by orders of magnitude beyond current rates in order for a direct search to perform better than a modulation search, suggesting that modulation searches are currently the best way to probe DM masses in the range 0.5 MeV $<$ $m_\chi$ $<$ 10 MeV, at least for the dark-photon mediator model considered in our work.

Experiments with noble-liquid targets have backgrounds across several electron bins, but are also able to take advantage of significantly larger exposures when compared with experiments using silicon. For a light dark-photon mediator, we find that experiments that use noble-liquid target detectors are unable to probe new parameter space for exposures as large as 1 tonne-year. However, we see that exposures of a tonne-month are sufficient to probe new parameter space for DM with masses $\sim$10~MeV $<$ $m_\chi$ $<$ 50~MeV in the case of a heavy dark-photon mediator and for the 1, 2, and 3\NE\ bins.

We make available on GitHub \url{https://github.com/ande8412/DarkMatterRates} the code to calculate dark-matter scattering rates with at different isoangles using the velocity distributions from {\tt Verne} and {\tt DaMaSCUS} as well as the code to generate our figures. We also make public our calculated velocity distribution data and modulation rates on Dryad at \url{https://datadryad.org/dataset/doi:10.5061/dryad.8pk0p2p19}.

\section*{Acknowledgments}
RE acknowledges support from DOE Grants DE-SC0017938 and DE-SC0025309, and Simons Investigator in Physics Awards~623940 and MPS-SIP-00010469, and Heising-Simons Foundation Grant No.~2017-380. 
XB initially received support from ANPCyT grant PICT-2021-I-A-01189. However, due to the discontinuation of funding by the Argentine government in a broader context where scientific research is facing significant challenges, he was compelled to leave Argentina and complete this work in France at IN2P3/CNRS.
TV is supported, in part, by the Israel Science Foundation (grant No. 1862/21), and by the NSF-BSF (grant No. 2021780).
RE and TV acknowledge support from the Binational Science Foundation (grant No.\ 2020220). T-TY is supported in part by NSF CAREER grant PHY-1944826. AD is supported by the Heising-Simons Foundation Grant No.~2017-380 through a subaward from Stony Brook University to the University of Oregon, and a Universities Research Associations fellowship. T-TY thanks the Università Degli Studi di Padova and the theory department at CERN for their hospitality and support where part of this work was completed. 

\appendix

\section{Discovery Potential in Noble Liquid Experiments}\label{app:noble}

In this appendix, we show the expected significance under specific background assumptions, which we list in Table~\ref{table:backgroundrates}.

\begin{table*}[!t]
\centering
\renewcommand{\arraystretch}{1.25}
\begin{tabular}{ |p{3cm}||p{3cm}|p{3cm}|p{3cm}|  }
 \hline
 \multicolumn{3}{|c|}{Assumed background rates for sensitivity projections [events/kg/day]} \\
 \hline
 \NE\ bin &  xenon & argon\\
 \hline
 $1e^-$   &  3  &   --\\
 $2e^-$   &  0.1  &  0.4\\
 $3e^-$   &  0.02  &  0.02\\
 $4e^-$   &  0.01  &  0.004\\
 \hline
\end{tabular}
\caption{Estimated background events in events/kg/day used in the significance studies. The background rates for the $2e^-$, $3e^-$ and $4e^-$ bins for argon were taken from the modeling in Figure 2 of~\cite{darkside2024}. For xenon we took the event rates from the D2 data set from Fig. 4 of~\cite{XENON:2024znc} and assumed they were all background events.}
\label{table:backgroundrates}
\end{table*}

\subsection{Xenon}
\begin{figure*}[!htb]
    \includegraphics[width=0.99\textwidth]{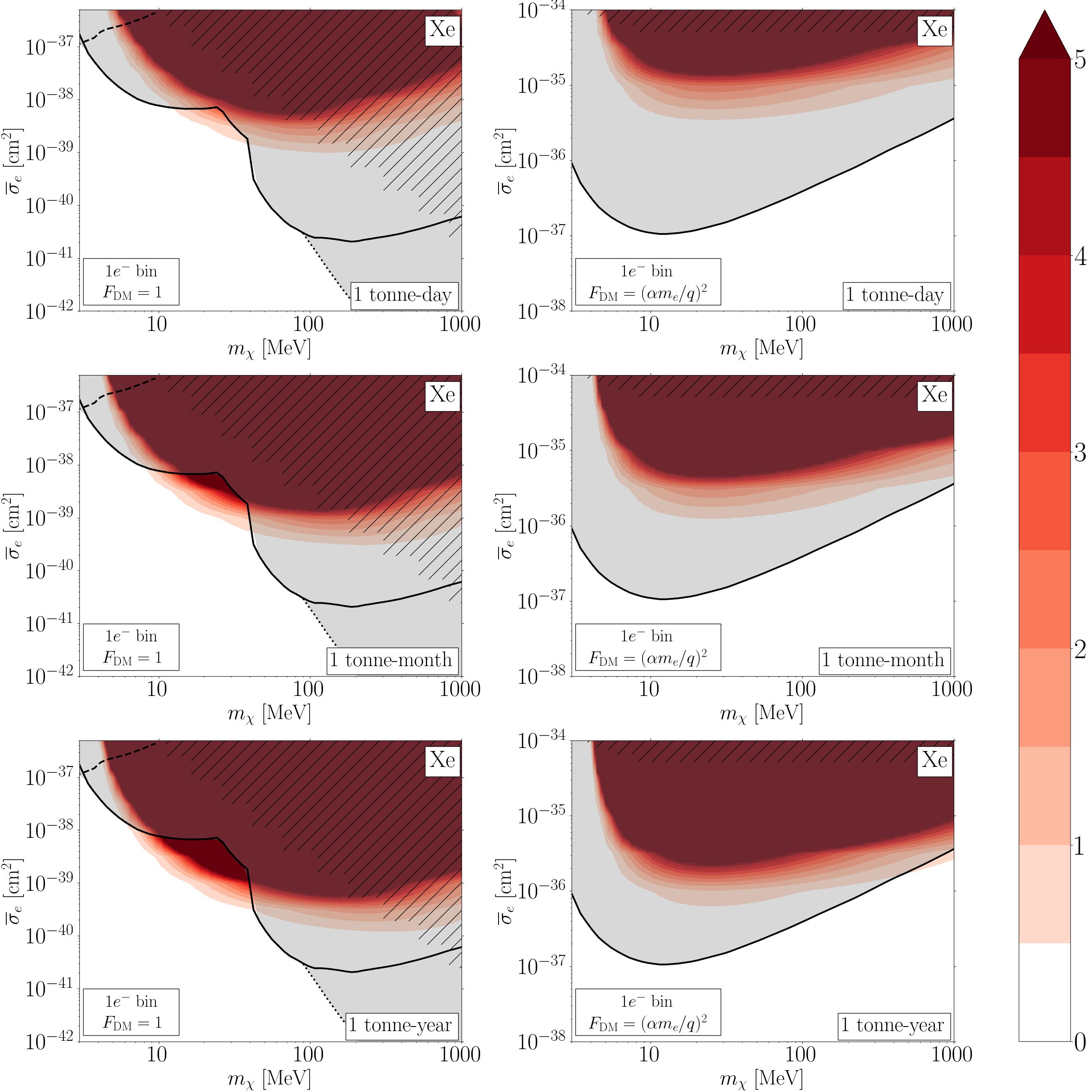}
    \caption{Expected significance contours from a modulation search in xenon for an exposure of 1 kg-day ({\bf row 1}), 1 kg-month ({\bf row 2}), and 1 tonne-month ({\bf row 3}) for the 1\NE\ bin for a detector located at SNOLAB, Canada. The background 1\NE\ rate is taken from~\cite{XENON:2024znc} assuming that the observed 1\NE\ rate consists entirely of background events and is listed in Table~\ref{table:backgroundrates}. The ({\bf left}) column is for a heavy dark-photon mediator, and the ({\bf right}) column is for a light dark-photon mediator. 
    Halo constraints (shaded region above solid black line) are combined using~\cite{damicmcollaboration2025probingbenchmarkmodelshiddensector,DamicModArnquist_2024,SENSEI:2024yyt,PandaXTLi_2023,XENON:2024znc,DarkSide:2022knj}. The {\bf black, dotted} curve shows constraints from the Migdal effect from~\cite{PandaXTLi_2023}.  Solar reflected DM constraints (shaded region above the black dashed line) come from~\cite{PhysRevLett.123.251801,Emken:2024nox,XENON:2024znc}. These significance contours are calculated using {\tt Verne} due to simulation limitations in {\tt DaMaSCUS}.  The {\bf hashed} region is where the mean free path of the DM through the Earth is less than $R_\oplus$, where the approximations used in {\tt Verne} break down.
    }
    \label{fig:Xenon1eSignificanceSnolab}
\end{figure*}

\begin{figure*}[!htb]
    \includegraphics[width=0.99\textwidth]{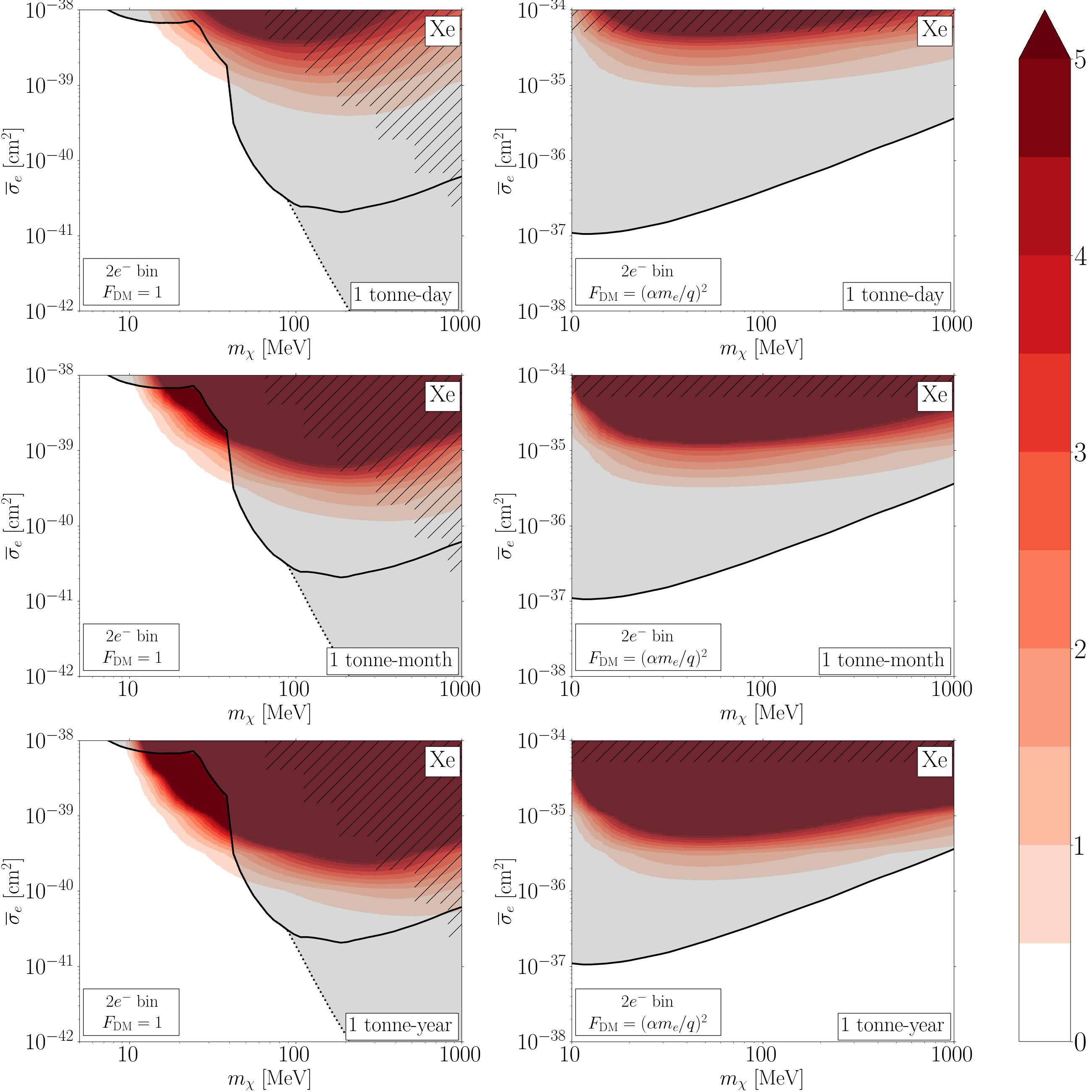}
    \caption{Same as Fig.~\ref{fig:Xenon1eSignificanceSnolab} but for the 2\NE\ bin.}
    \label{fig:Xenon2eSignificanceSnolab}
\end{figure*}

\begin{figure*}[!htb]
    \includegraphics[width=0.99\textwidth]{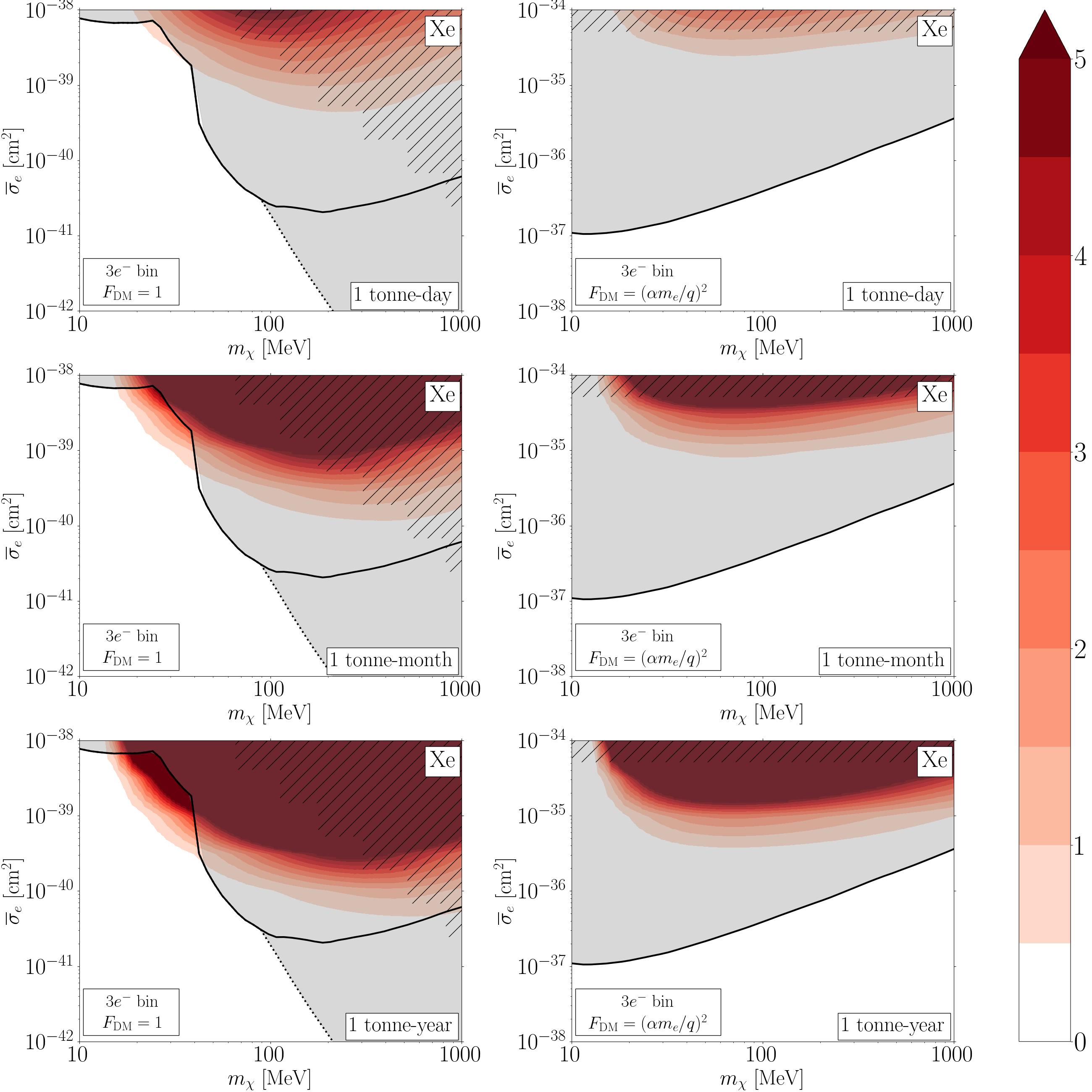}
    \caption{Same as Fig.~\ref{fig:Xenon1eSignificanceSnolab} but for the 3\NE\ bin.}
    \label{fig:Xenon3eSignificanceSnolab}
\end{figure*}

\begin{figure*}[!htb]
    \includegraphics[width=0.99\textwidth]{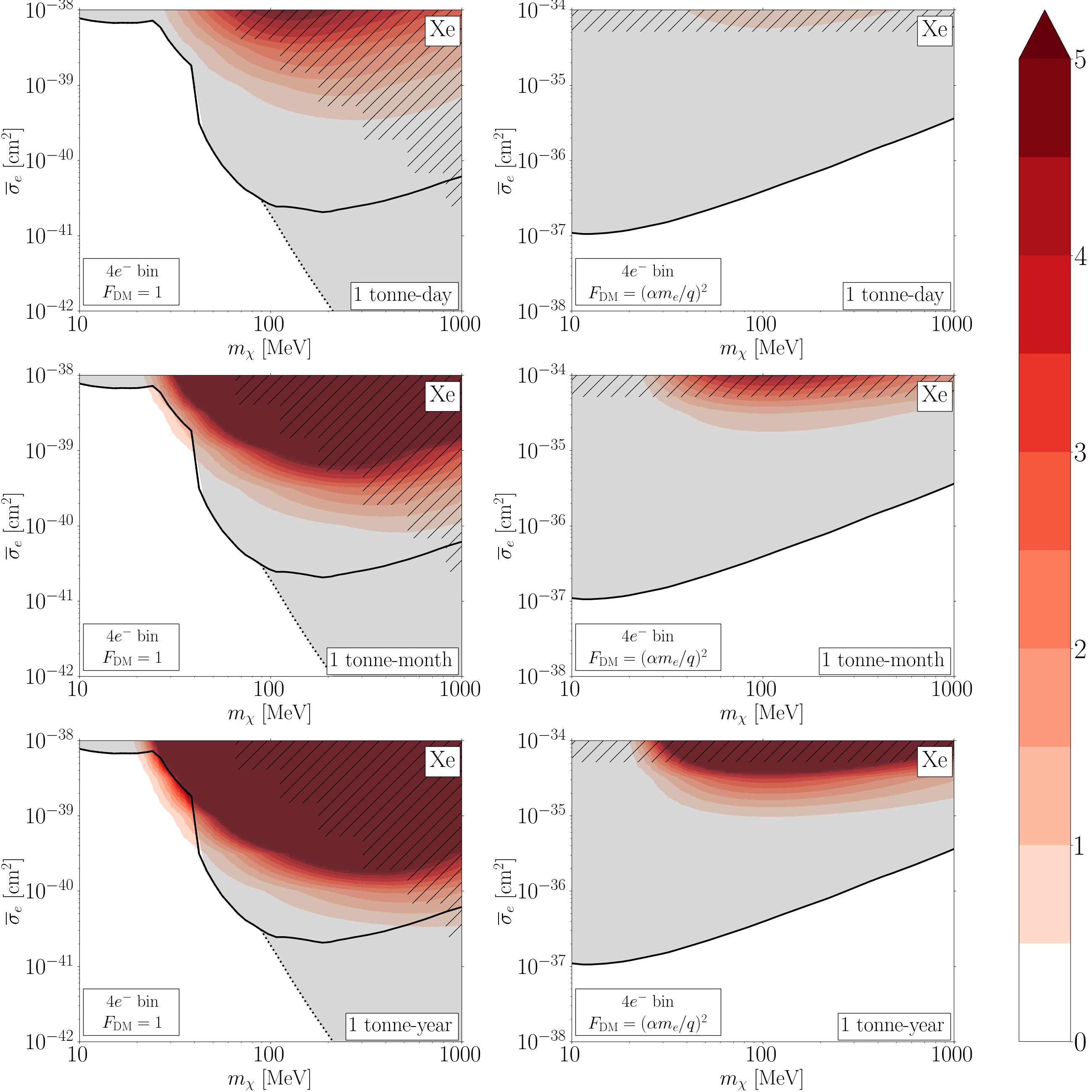}
    \caption{Same as Fig.~\ref{fig:Xenon1eSignificanceSnolab} but for the 4\NE\ bin.}
    \label{fig:Xenon4eSignificanceSnolab}
\end{figure*}

To determine the discovery potential using a modulation search in liquid xenon experiments, we take events from the D2 data set in Fig.~4 of~\cite{XENON:2024znc} and assume these are all background events (summarized in Table~\ref{table:backgroundrates}). We then calculate the expected signal $\Delta S$ using a modified version of {\tt wimprates} and the results from {\tt Verne} at three exposures: 1 tonne-day, 1 tonne-month, and 1 tonne-year. Significance contours for the heavy dark-photon mediator are plotted in Figs.~\ref{fig:Xenon1eSignificanceSnolab},~\ref{fig:Xenon2eSignificanceSnolab},~\ref{fig:Xenon3eSignificanceSnolab}, and~\ref{fig:Xenon4eSignificanceSnolab}. For the heavy dark-photon mediator, we see that constraints can be improved from $\approx$10~MeV--50~MeV, especially using the 1\NE\, 2\NE\ and 3\NE\ bins. The regions accessible with the 4\NE\ bin have already been excluded. The light dark-photon mediator is more challenging to probe as the fractional modulation flattens (an effect that can be seen in Fig.~\ref{fig:mfp}) and is small near current constraints. 


\subsection{Argon}
\begin{figure*}[!htb]
    \includegraphics[width=0.99\textwidth]{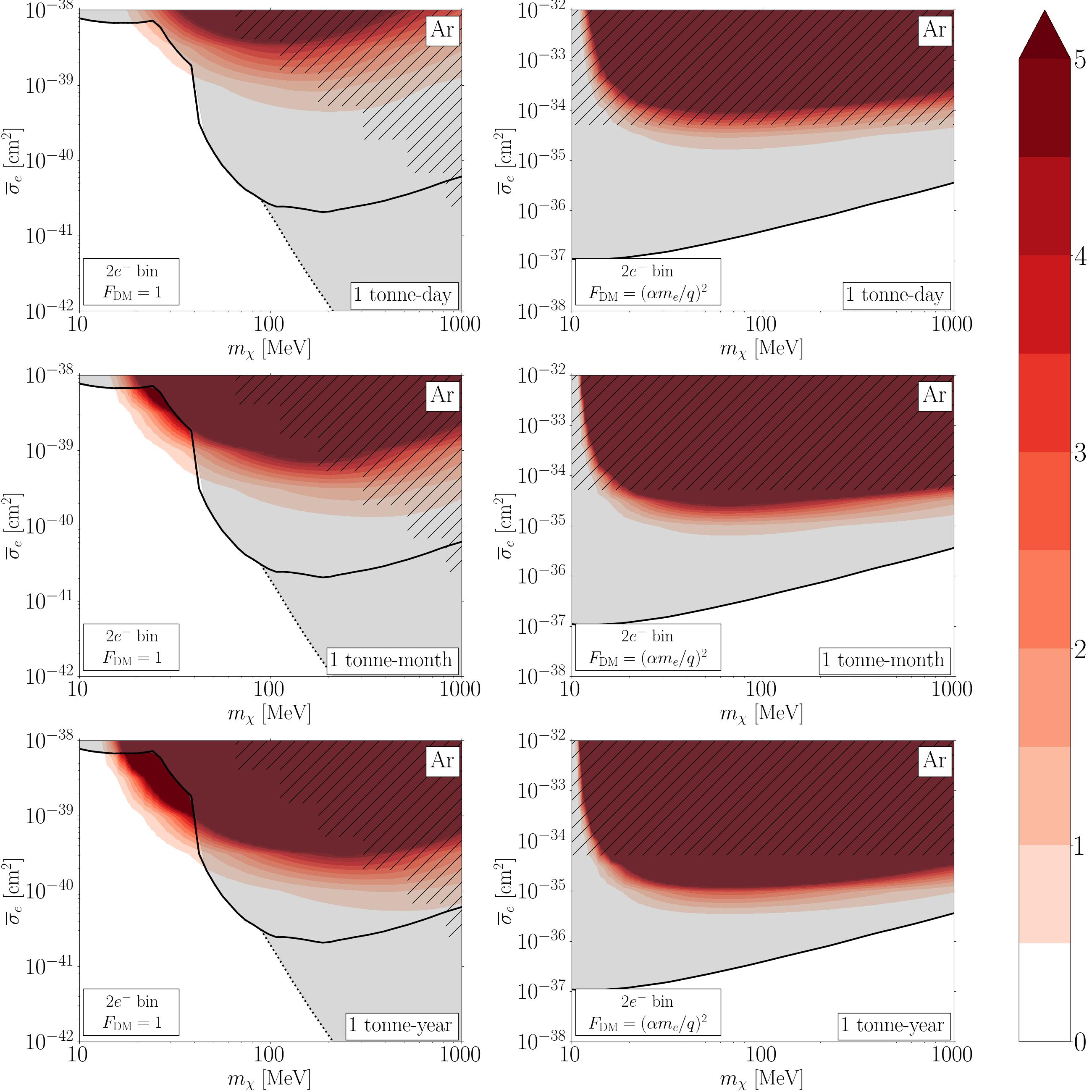}
   \caption{Expected significance contours from a modulation search in argon for an exposure of 1 kg-day ({\bf row 1}), 1 kg-month ({\bf row 2}), and 1 tonne-month ({\bf row 3}) for the 1\NE\ bin for a detector located at SNOLAB, Canada. The background 2\NE\ rate is taken from figure 2 in~\cite{darkside2024} assuming the modeled background rate holds and is summarized in Table~\ref{table:backgroundrates}. The ({\bf left}) column is for a heavy dark-photon mediator, and the ({\bf right}) column is for a light dark-photon mediator. 
    Halo constraints (shaded region above solid black line) are combined using~\cite{damicmcollaboration2025probingbenchmarkmodelshiddensector,DamicModArnquist_2024,SENSEI:2024yyt,PandaXTLi_2023,XENON:2024znc,DarkSide:2022knj}. The {\bf black, dotted} curve shows constraints from the Migdal effect from~\cite{PandaXTLi_2023}.  Solar reflected DM constraints (shaded region above the black dashed line) come from~\cite{PhysRevLett.123.251801,Emken:2024nox,XENON:2024znc}. These significance contours are calculated using {\tt Verne} due to simulation limitations in {\tt DaMaSCUS}.  The {\bf hashed} region is where the mean free path of the DM through the Earth is less than $R_\oplus$, where the approximations used in {\tt Verne} break down.}
    \label{fig:Argon2eSignificanceSnolab}
\end{figure*}

\begin{figure*}[!htb]
    \includegraphics[width=0.99\textwidth]{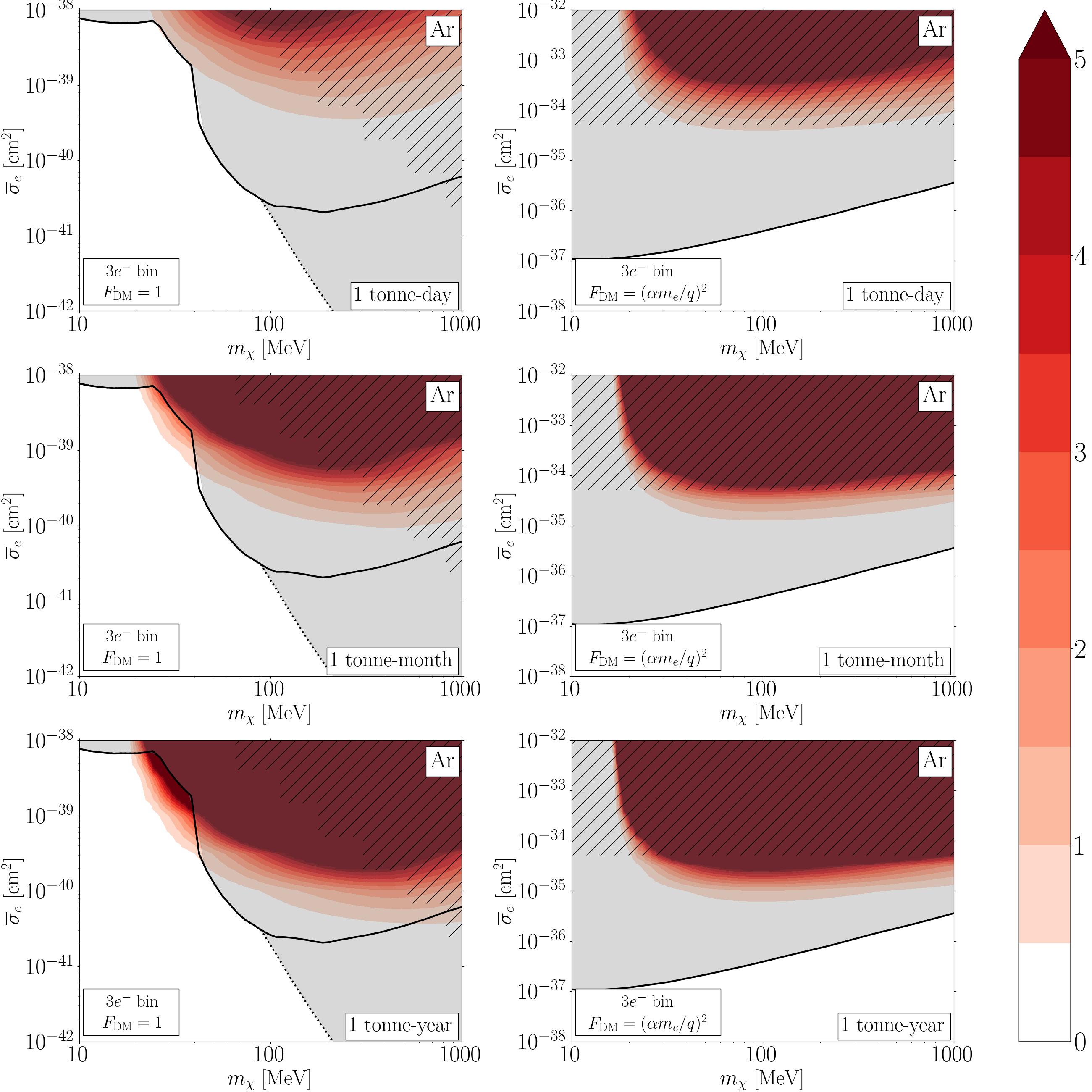}
    \caption{Same as Fig.~\ref{fig:Argon2eSignificanceSnolab} but for the 3\NE\ bin.}
    \label{fig:Argon3eSignificanceSnolab}
\end{figure*}

\begin{figure*}[!htb]
    \includegraphics[width=0.99\textwidth]{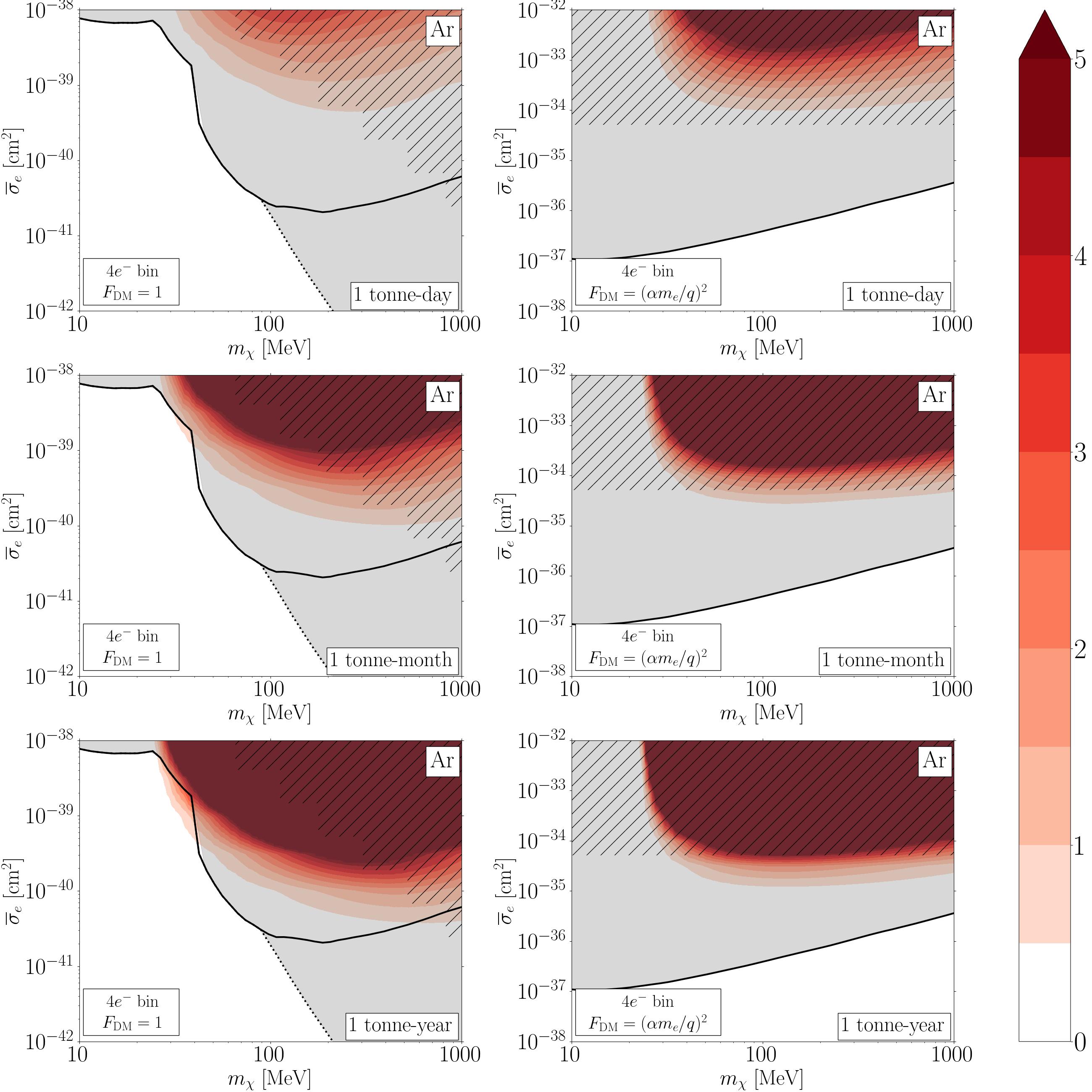}
    \caption{Same as Fig.~\ref{fig:Argon2eSignificanceSnolab} but for the 4\NE\ bin.}
\label{fig:Argon4eSignificanceSnolab}
\end{figure*}

To determine the discovery potential using a modulation search in liquid argon experiments, we use the expected modeled background rate from Fig.~2 of~\cite{darkside2024} for the $2e^-$, $3e^-$ and $4e^-$ bins and calculate the expected signal $\Delta S$ using a modified version of {\tt wimprates} as discussed in~\ref{sec:mod} and the results from {\tt Verne}  at three exposures: 1 tonne-day, 1 tonne-month, and 1 tonne-year. Significance contours for the heavy mediator are plotted in Figs.~\ref{fig:Argon2eSignificanceSnolab}, ~\ref{fig:Argon3eSignificanceSnolab}, and~\ref{fig:Argon4eSignificanceSnolab}.  While we could not test the 1\NE\ bin due to a lack of information about the background rate, we see that constraints can be improved in a similar mass range to xenon for both the 2\NE\ and 3\NE\ bins for the heavy mediator case.  Similarly to xenon, the 4\NE\ bin turns out to be less useful. For the same reasons as for the xenon targets, the light dark-photon mediator remains out of reach to probe with daily modulation. 

\section{Sensitivity Figures for the Southern Hemisphere}\label{app:south}

\begin{figure*}[!htb]
    \includegraphics[width=0.99\textwidth]{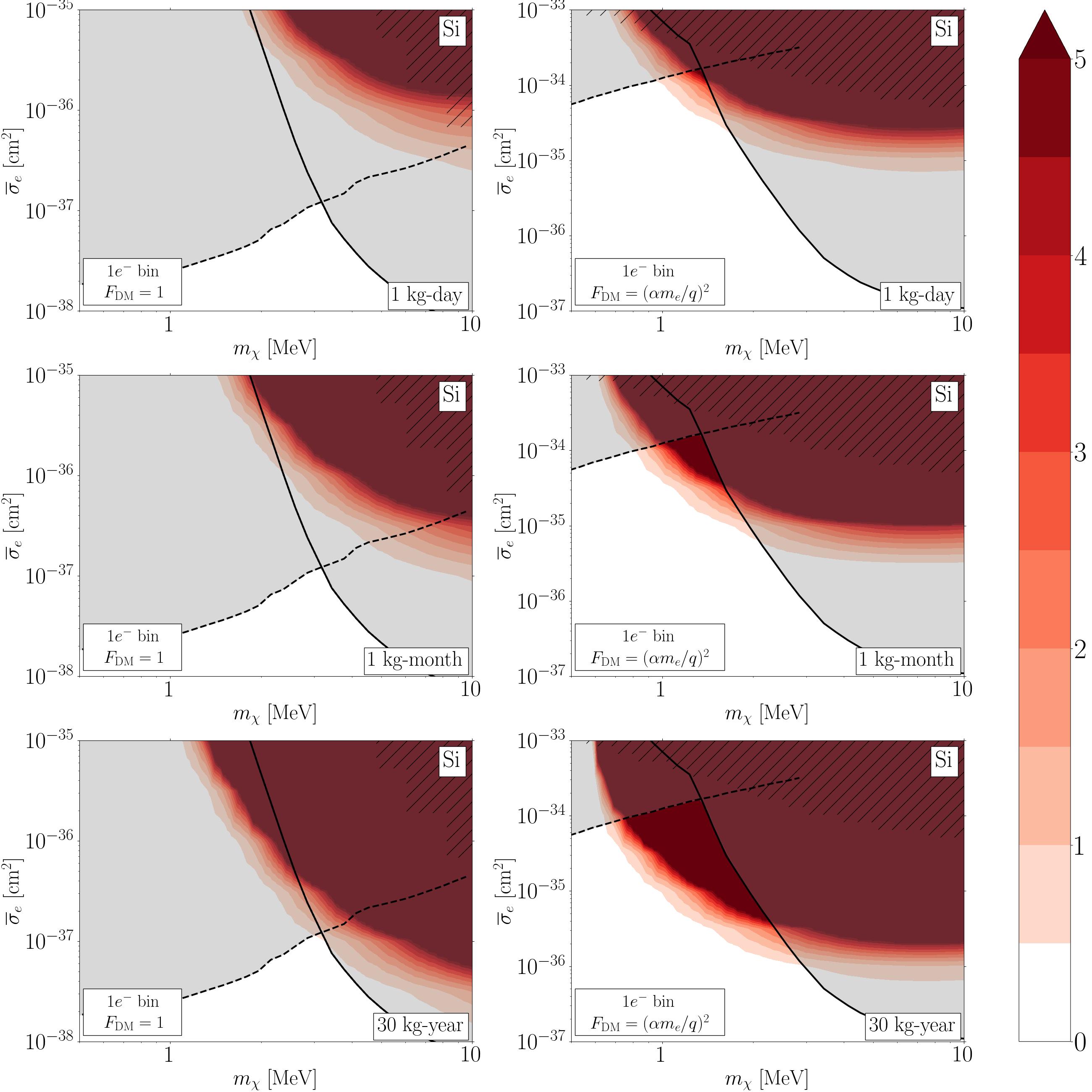}
    \caption{Same as Fig.~\ref{fig:Silicon1eSignificanceSnolab} but for a detector located at Stawell, Australia.}
    \label{fig:Silicon1eSignificanceSUPL}
\end{figure*}

\begin{figure*}
    \includegraphics[width=0.99\textwidth]{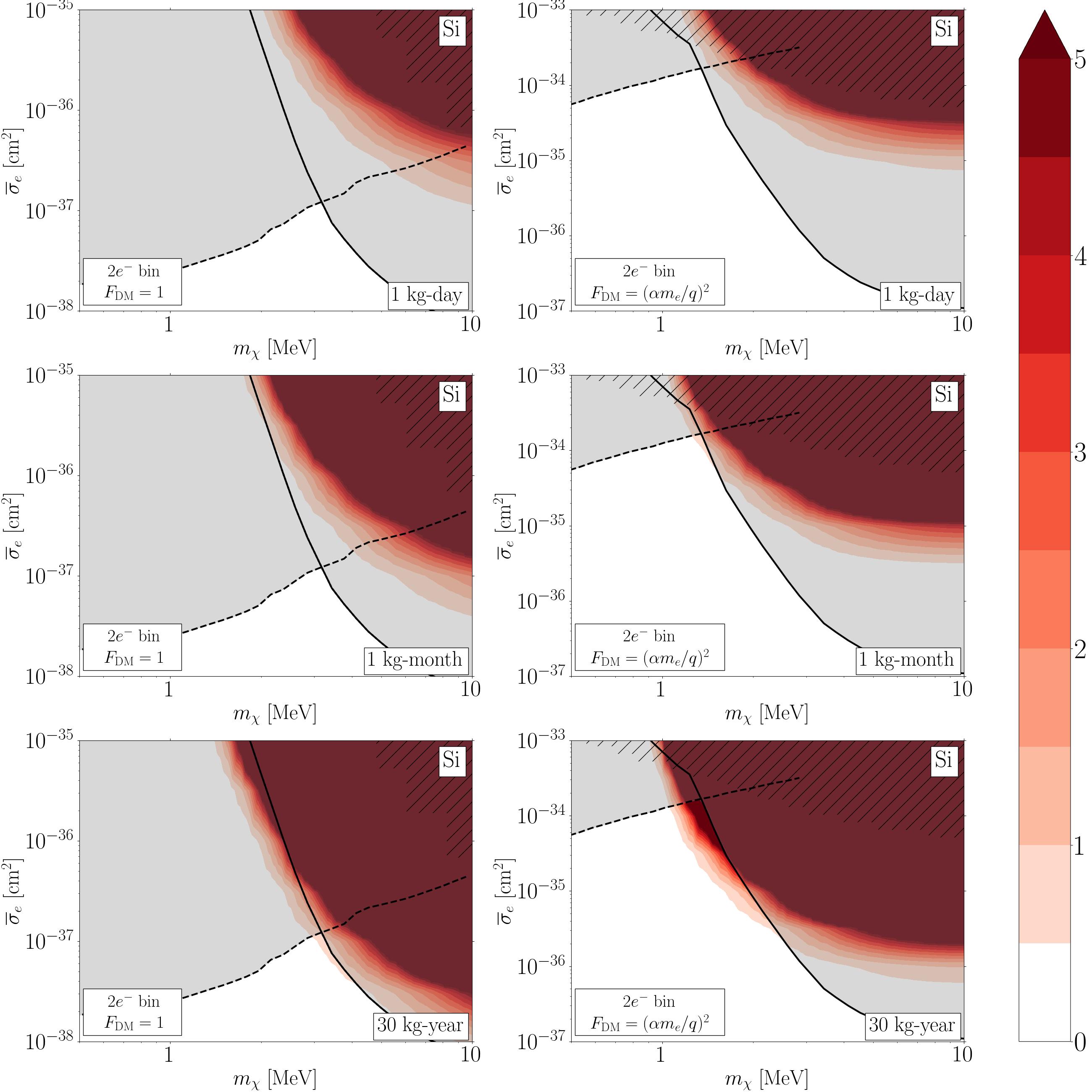}
    \caption{Same as Fig.~\ref{fig:Silicon2eSignificanceSnolab} but for a detector located at Stawell, Australia.}
    \label{fig:Silicon2eSignificanceSUPL}
\end{figure*}

\begin{figure*}
    \includegraphics[width=0.99\textwidth]{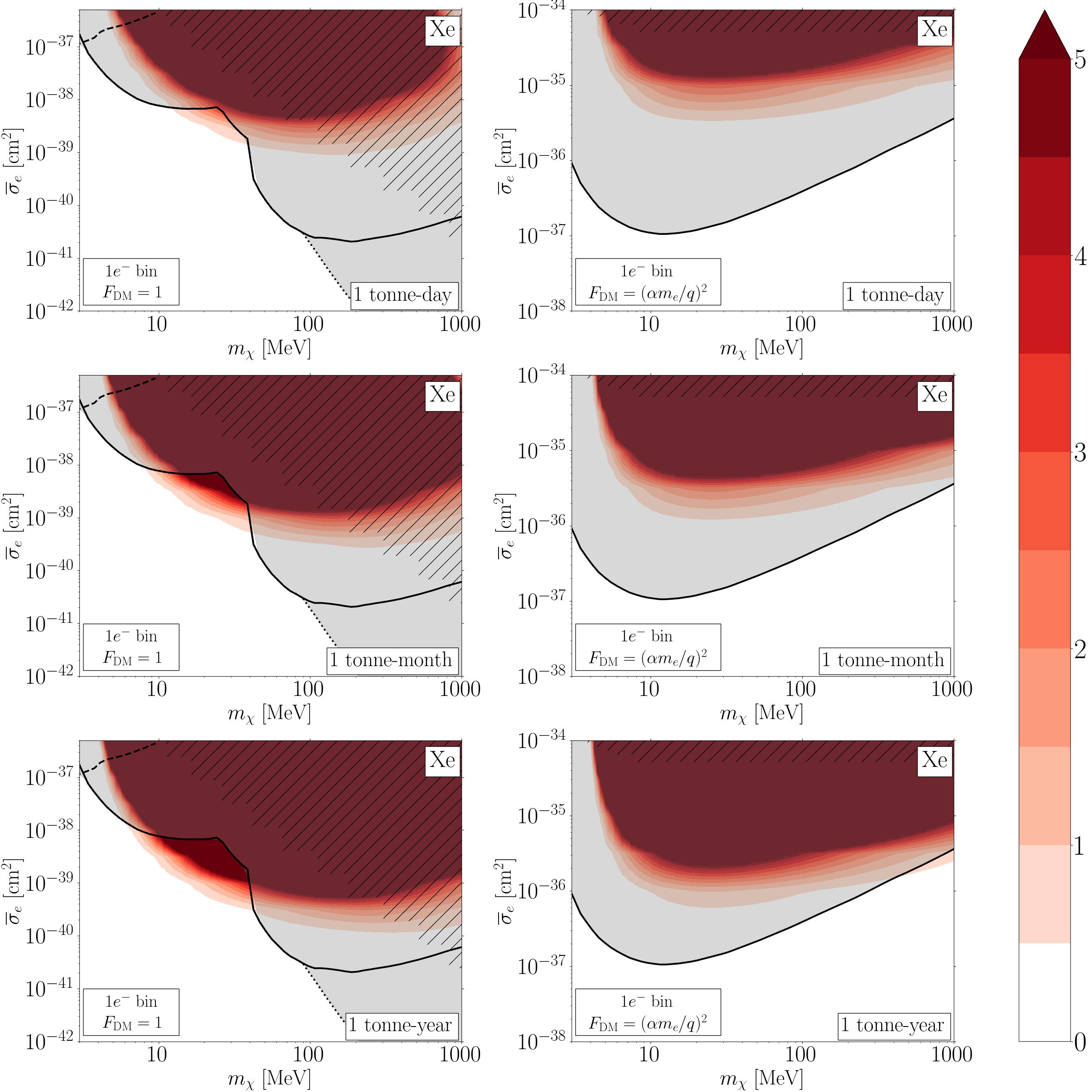}
    \caption{Same as Fig.~\ref{fig:Xenon1eSignificanceSnolab} but for a detector located at Stawell, Australia.}
    \label{fig:Xenon1eSignificanceSUPL}
\end{figure*}

\begin{figure*}
    \includegraphics[width=0.99\textwidth]{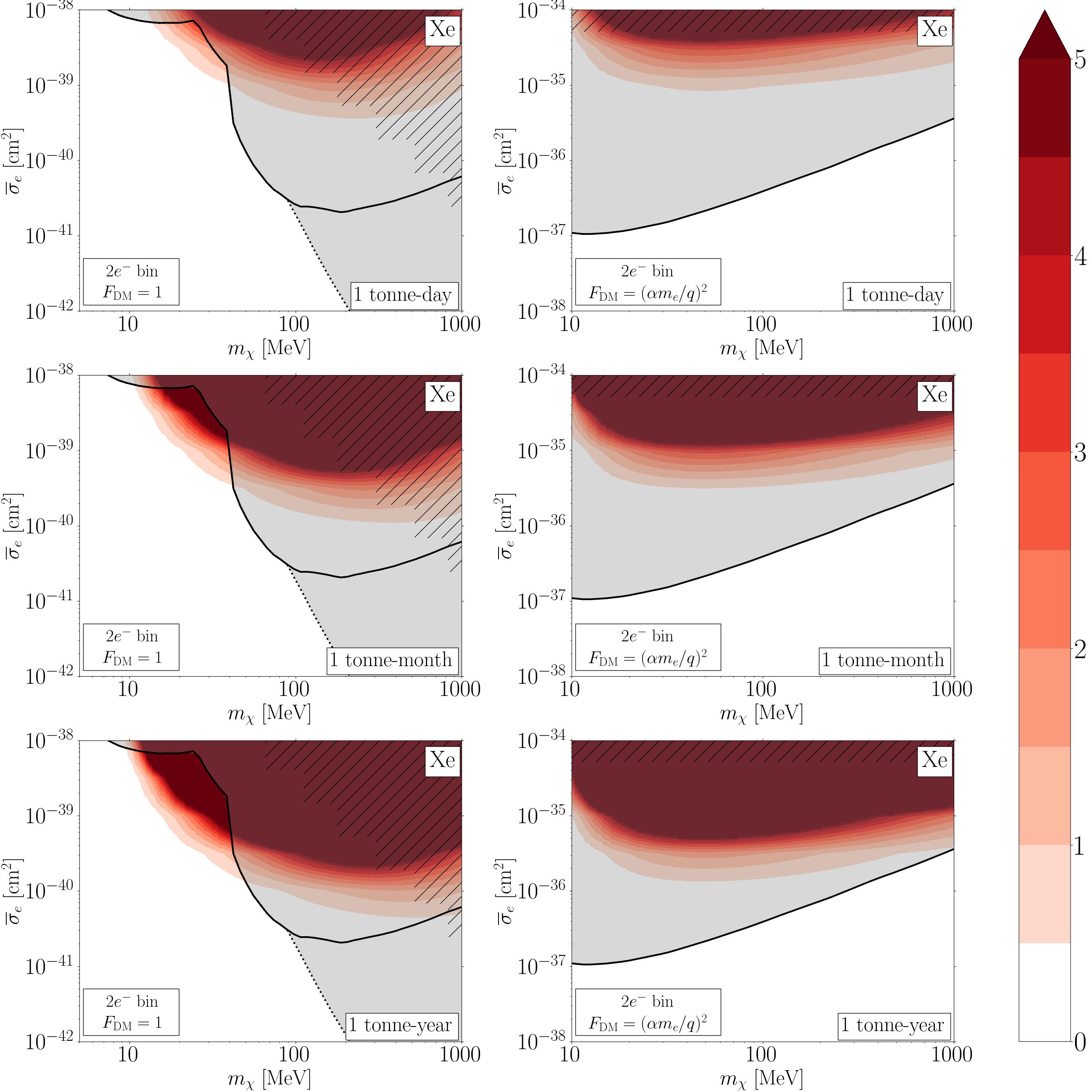}
    \caption{Same as Fig.~\ref{fig:Xenon2eSignificanceSnolab} but for a detector located at Stawell, Australia.}
    \label{fig:Xenon2eSignificanceSUPL}
\end{figure*}

\begin{figure*}
    \includegraphics[width=0.99\textwidth]{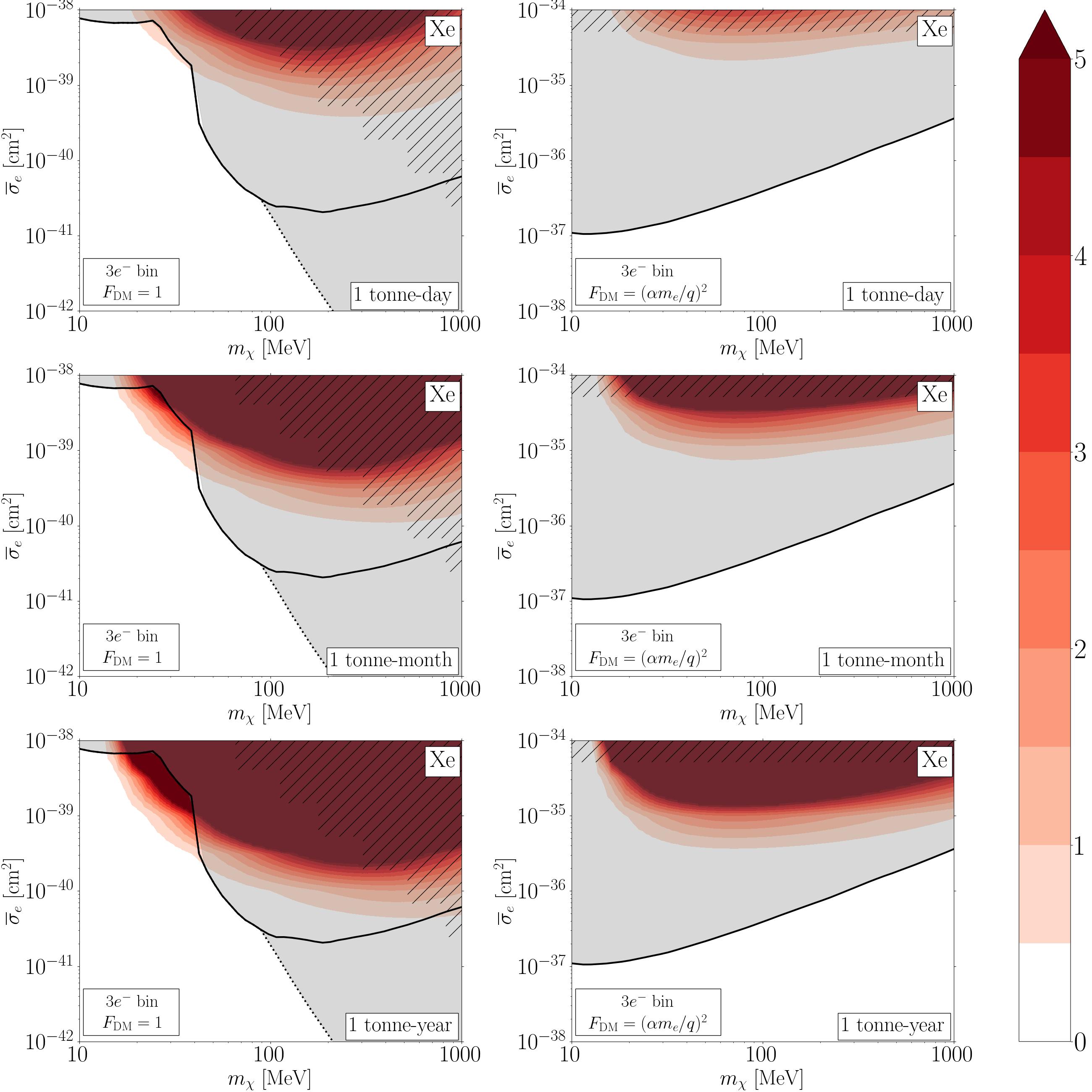}
    \caption{Same as Fig.~\ref{fig:Xenon3eSignificanceSnolab} but for a detector located at Stawell, Australia.}
    \label{fig:Xenon3eSignificanceSUPL}
\end{figure*}

\begin{figure*}
    \includegraphics[width=0.99\textwidth]{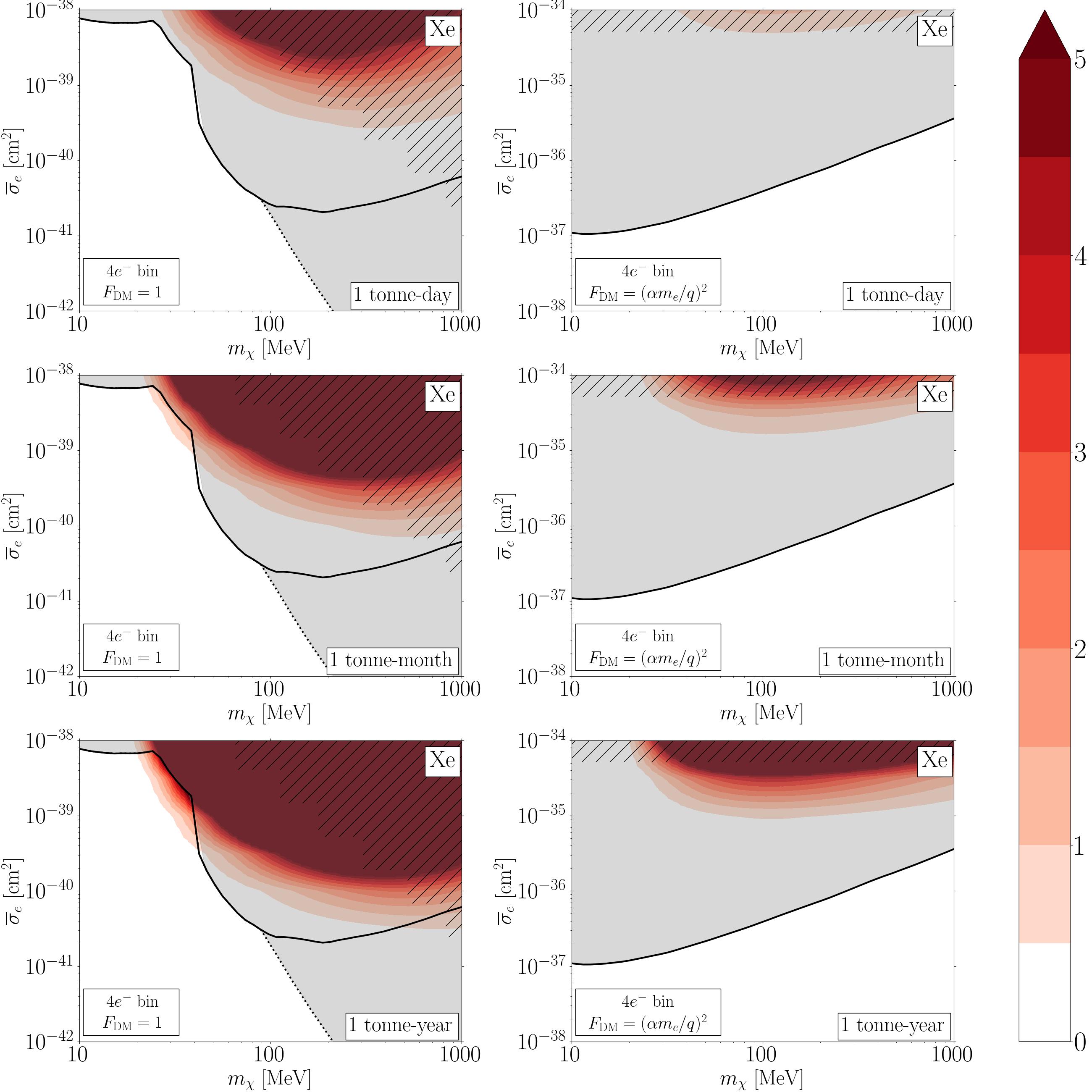}
    \caption{Same as Fig.~\ref{fig:Xenon4eSignificanceSnolab} but for a detector located at Stawell, Australia.}
    \label{fig:Xenon4eSignificanceSUPL}
\end{figure*}

\begin{figure*}
    \includegraphics[width=0.99\textwidth]{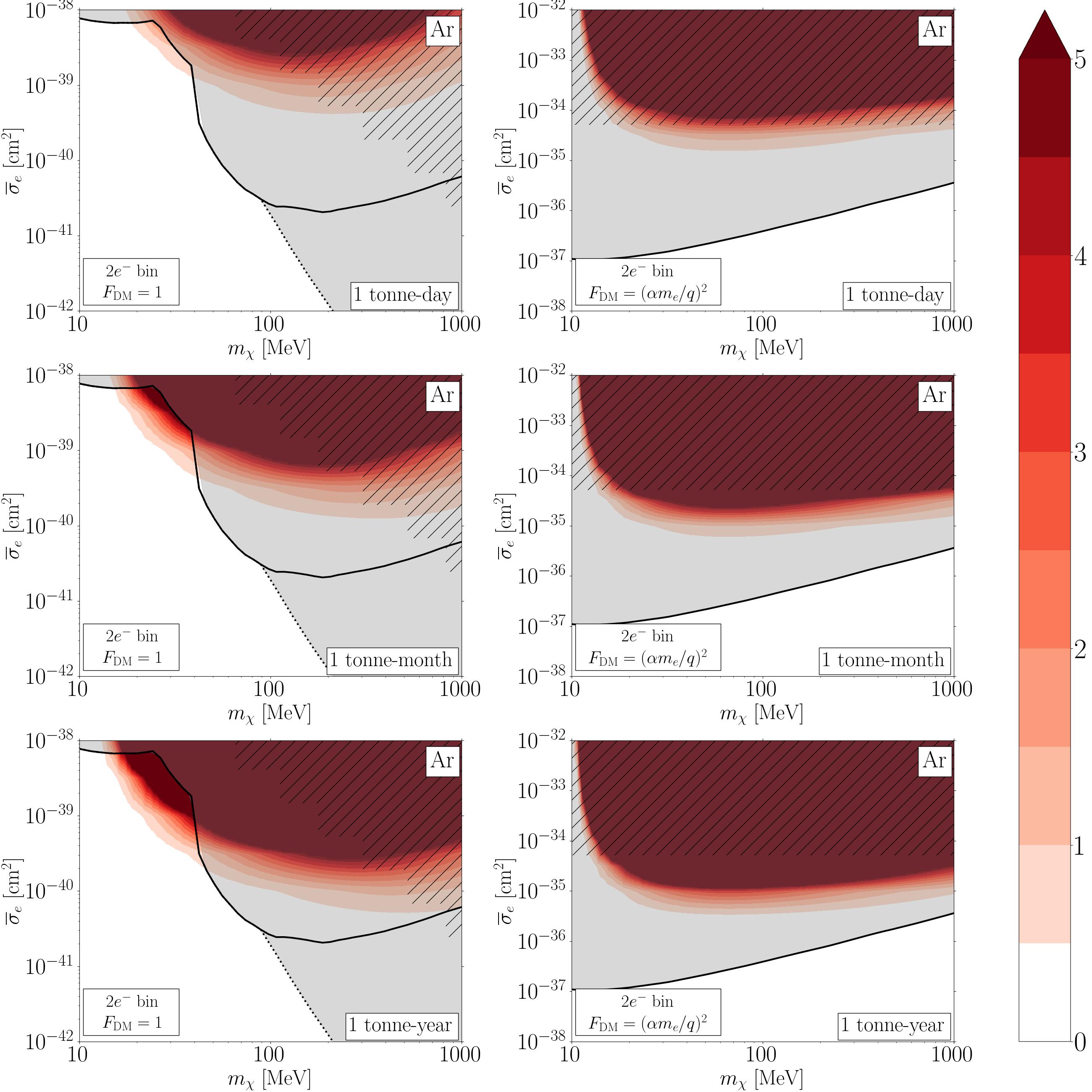}
    \caption{Same as Fig.~\ref{fig:Argon2eSignificanceSnolab} but for a detector located at Stawell, Australia.}
    \label{fig:Argon2eSignificanceSUPL}
\end{figure*}

\begin{figure*}
    \includegraphics[width=0.99\textwidth]{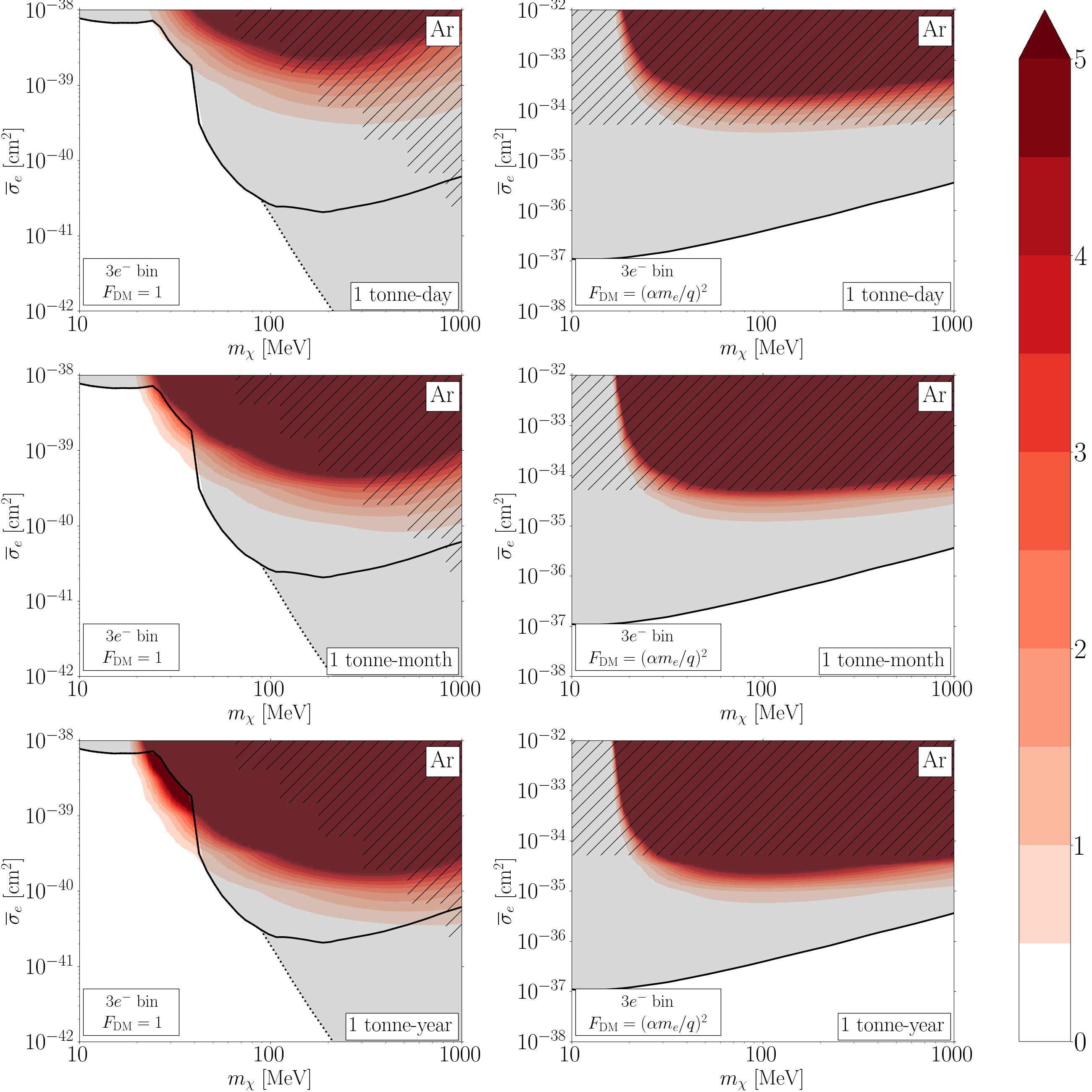}
    \caption{Same as Fig.~\ref{fig:Argon3eSignificanceSnolab} but for a detector located at Stawell, Australia.}
    \label{fig:Argon3eSignificanceSUPL}
\end{figure*}

\begin{figure*}
    \includegraphics[width=0.99\textwidth]{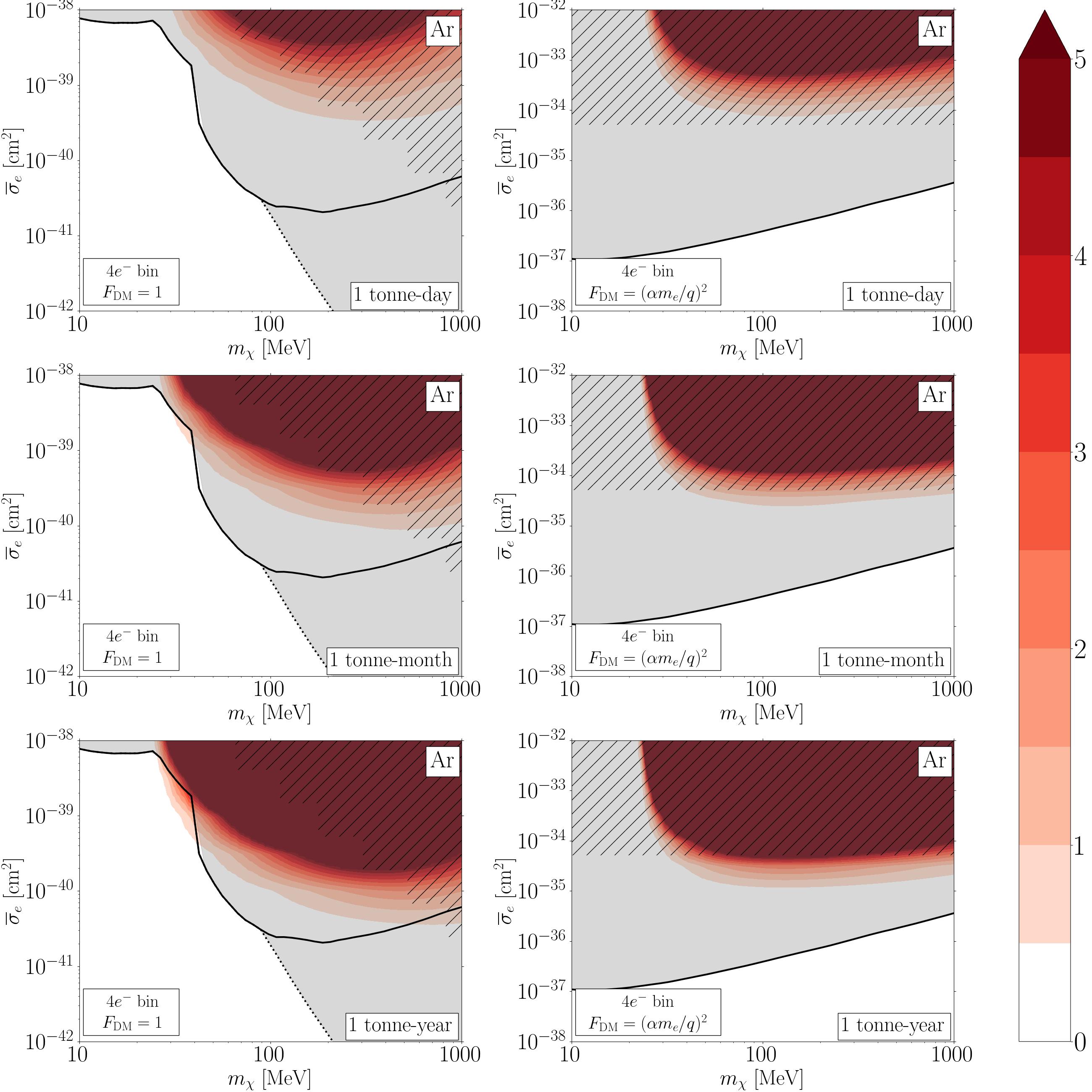}
    \caption{Same as Fig.~\ref{fig:Argon4eSignificanceSnolab} but for a detector located at Stawell, Australia.}
    \label{fig:Argon4eSignificanceSUPL}
\end{figure*}

Here we reproduce discovery potential figures in silicon, xenon and argon for our representative location (Stawell, Australia) for the Southern Hemisphere in 
Figs.~\ref{fig:Silicon1eSignificanceSUPL},~\ref{fig:Silicon2eSignificanceSUPL},~\ref{fig:Xenon1eSignificanceSUPL},~\ref{fig:Xenon2eSignificanceSUPL},~\ref{fig:Xenon3eSignificanceSUPL},~\ref{fig:Xenon4eSignificanceSUPL},~\ref{fig:Argon2eSignificanceSUPL},~\ref{fig:Argon3eSignificanceSUPL}, and~\ref{fig:Argon4eSignificanceSUPL}. 
The fractional modulation is significantly greater in the Southern Hemisphere in the parts of the parameter space that are already excluded (i.e., for larger cross-sections); however, for the parameter regions close to the current bounds, the fractional modulation is similar, and hence the expected sensitivities are also similar, as can be see when comparing these plots to those in Figs.~\ref{fig:Fractional Modulation Amplitude FDM1} and~\ref{fig:Fractional Modulation Amplitude FDMq2}. 

\newpage

\newpage
\newpage
\newpage
\newpage
\newpage


\clearpage
\bibliographystyle{JHEP}
\bibliography{references.bib}

\end{document}